\documentclass[submission,Phys]{SciPost}
\usepackage{amsfonts,amssymb,amsmath}
\usepackage{mathrsfs}
\usepackage{mathtools}
\usepackage[utf8]{inputenc}
\usepackage{graphicx}
\usepackage{hyperref}
\usepackage{filecontents}
\usepackage{float}

\newcommand{\ket}[1]{\left|#1\right>}
\newcommand{\bra}[1]{\left<#1\right|}

\begin{document}

\begin{center}{\Large \textbf{
Domain wall problem in the quantum XXZ chain and semiclassical behavior close to the isotropic point
}}\end{center}

\begin{center}
G. Misguich\textsuperscript{1,2*},
N. Pavloff\textsuperscript{3},
V. Pasquier\textsuperscript{1}
\end{center}
\begin{center} {\bf 1} Institut de Physique Théorique, Université
  Paris Saclay, CEA, CNRS UMR 3681, 91191 Gif-sur-Yvette, France
  \\
  {\bf 2} Laboratoire de Physique Théorique et Modélisation, CNRS UMR
  8089, Université de Cergy-Pontoise, 95302 Cergy-Pontoise, France
  \\
  {\bf 3} Laboratoire de Physique Théorique et Modèles Statistiques,
  CNRS UMR 8626, Univ. Paris-Sud, Université Paris Saclay, 91405 Orsay,
  France
  \\
  * gregoire.misguich@cea.fr
\end{center}

\begin{center}
\today
\end{center}
\section*{Abstract} {\bf We study the dynamics of a spin-$\frac{1}{2}$
  XXZ chain which is initially prepared in a domain-wall state.  We
  compare the results of time-dependent Density Matrix Renormalization
  Group simulations with those of an effective description in terms of
  a classical anisotropic Landau-Lifshitz (LL) equation.  Numerous
  quantities are analyzed: magnetization ($x$, $y$ and $z$ components),
  energy density, energy current, but also some spin-spin correlation
  functions or entanglement entropy in the quantum chain.  Without any
  adjustable parameter a quantitative agreement is observed between
  the quantum and the LL problems in the long time limit, when the
  models are close to the isotropic point. This is explained as a consequence
  of energy conservation.  At the isotropic point the mapping between
  the LL equation and the nonlinear Schrödinger equation is used to
  construct a variational solution capturing several aspects of the
  problem.}
\vspace{10pt}
\noindent\rule{\textwidth}{1pt}
\tableofcontents\thispagestyle{fancy}
\noindent\rule{\textwidth}{1pt}
\vspace{10pt}

\section{Introduction}

The study of out-of-equilibrium quantum systems constitutes a major
research field in condensed matter and quantum statistical physics.
The understanding of the dynamics of isolated quantum many-body
systems after a quench is one of the important questions in
this domain~\cite{polkovnikov_colloquium_2011,eisert_quantum_2015}.
It addresses the long-time behavior of the system after a sudden
change of some external parameter at time $t=0$, and, for instance,
the possible relaxation of local observables to steady values, and the
characterization of the associated steady state.

In the field of quantum quenches, the situation of integrable systems
is very peculiar~\cite{calabrese_introduction_2016}. Their dynamics
is qualitatively different from that of generic models where the only
local conserved quantities are the energy and the momentum. Indeed,
in integrable systems the large number of local conserved quantities 
represents an enhanced memory of the initial state. In situations
where a non-integrable system would approach a Gibbs state
characterized by a single quantity -- the temperature or the energy --
an integrable system will keep a much more detailed memory of its
initial conditions. When the system is spatially homogeneous, this can
give rise to so-called generalized Gibbs
ensembles~\cite{rigol_relaxation_2007}. This also leads to unconventional
transport properties (see, e.g., Ref. \cite{castella_integrability_1995,zotos_transport_1997,ilievski_microscopic_2017}).

In the case of an inhomogeneous quench, the system is initially
prepared at $t=0$ in a state which is spatially inhomogeneous, while
the Hamiltonian driving the unitary evolution at $t>0$ can be taken to
be translation invariant.  As an example one may suddenly join at
$t=0$ two macroscopically different and homogeneous states, having,
for instance, different particle densities, different magnetizations
or different temperatures.

Such protocols are interesting because they can produce
current-carrying states, and they allow to address questions about
transport (ballistic or diffusive?) or to test hydrodynamic
descriptions.  In integrable one-dimensional systems, like the
spin-$\frac{1}{2}$ XXZ chain or the Bose gas with $\delta$
interactions~\cite{lieb_exact_1963}, an important progress has been
made recently concerning the dynamics of inhomogeneous states.  An
hydrodynamic description taking into account the local conserved
quantities was developed, which is now called ``generalized
hydrodynamics" (GHD)
~\cite{castro-alvaredo_emergent_2016,bertini_transport_2016}.  It has
lead to many fruitful developments in the last few
years~\cite{piroli_transport_2017,ilievski_microscopic_2017,de_luca_nonequilibrium_2017,ilievski_ballistic_2017,bulchandani_classical_2017,fagotti_higher-order_2017,bulchandani_classical_2017,bulchandani_bethe-boltzmann_2018,doyon_geometric_2018,alba_entanglement_2018,collura_analytic_2018,de_nardis_hydrodynamic_2018,bastianello_generalized_2018},
including on the experimental side \cite{schemmer_generalized_2019}.

One of the applications of GHD is a classic inhomogeneous quench
problem
\cite{antal_transport_1999,gobert_real-time_2005,mossel_relaxation_2010},
where an XXZ spin chain is prepared at time $t=0$ in a domain-wall state
where all the spins in the left half of the chain are pointing ``up'',
and the spins in the right half are pointing ``down''.
This setup leads to three regimes, depending on the anisotropy
parameter $\Delta$ in the Hamiltonian of the chain.  For $|\Delta|<1$
(easy-plane) the dynamics creates a $z$-magnetization profile which
extends ballistically with time. $\Delta=0$ is a simple limit
(Jordan-Wigner mapping to free fermions) where such a ballistic
transport can be compared with exact
calculations~\cite{antal_transport_1999}.  For $|\Delta|>1$
(easy-axis) this profile gets frozen at long time
\cite{gobert_real-time_2005,mossel_relaxation_2010}.  While the GHD
can be used to study the system in the ballistic cases in details, it
does not predict much more than the absence of ballistic transport for
$|\Delta|\geq 1$. Numerical simulations however suggests a diffusive
behavior with logarithmic corrections at the isotropic point
$\Delta=1$~\cite{misguich_dynamics_2017}.

In a recent paper Gamayun {\it el al.}\cite{gamayun_domain_2019}
(we also mention \cite{mallick_2017}) considered the domain wall
problem in a different model, which is a classical ferromagnetic chain
in the continuum limit.  There, the magnetization $\vec M$ is a
classical unit vector which depends on time and on a continuous space
variable $r$, and its (precession) dynamics is described by the
celebrated anisotropic Landau Lifshitz (LL) equation. The initial
condition taken in Ref. \cite{gamayun_domain_2019} is a smooth function or
$r$ interpolating between $\vec M=\vec e_z$ at $r=-\infty$ to $\vec
M=-\vec e_z$ at $r=+\infty$, which is the continuum analog of the
domain wall in the context of lattice spin chains.  The main
result of \cite{gamayun_domain_2019} is that the dynamics of
$M^z(r,t)$ follows three regimes, and is qualitatively similar to the
quantum case: ballistic in the easy-plane regime, frozen in the
easy-axis, and diffusive with logarithmic correction in the isotropic
model.

In the present paper we present some quantitative comparisons
for the domain wall problem i) in the XXZ chain, and ii) in
the anisotropic LL system. We investigate numerous quantities: the
magnetization ($z$ component but also $x$ and $y$ components), the energy
density, the energy current, but also some spin-spin correlation
functions or the entanglement entropy in the quantum chain. We analyze
the ``diffusion'' region, of size $\sqrt{t}$ (with multiplicative
logarithmic corrections), but also the regime where $r/t$ is finite,
and where both the LL and XXZ problems show some nontrivial
behavior. These results are obtained using time-dependent
density-matrix renormalization group (tDMRG) simulations of the
spin-$\frac{1}{2}$ model, as well as numerical and analytical
calculations for the LL problem (hydrodynamic approximation, perturbative
expansion, or mapping to the nonlinear Schrödinger equation (NLS) and
variational ansatz).  As an important result, we observe that the
similarities between the quantum spin chain and the LL problem are not
only qualitative, but semi-quantitative (or even quantitative) close
to the isotropic point, in the long time limit, and without any
adjustable parameter. As explained in Sec.~\ref{sec:energy}, we argue
that this is a simple consequence of energy conservation.
Sec.~\ref{sec:easy-plane} deals with the easy-plane case. It is shown
that the LL problem gives a linear $z$-magnetization profile which is
identical to that of XXZ problem in the limit $\Delta\to 1^-$.  While
the presence of a linear profile in the long time limit was known for
these two models, we show that they have exactly the same slope in the
limit $\Delta\to 1^-$, and similar finite-time corrections.
In Sec.~\ref{sec:sw} we analyze the LL problem in the limit where the
magnetization is close to $M^z=\pm1$. This perturbative expansion turns
out to correctly describe the regime when $|r|/t$ is large for LL,
with small amplitude oscillations of $\vec M$ around the $z$ axis. But
it also describes some aspects of the XXZ problem, like the tail of
the $\left<S^z\right>$ or energy density profiles.  In
Sec.~\ref{sec:easy_axis} we discuss the spatial width of the stationary
profile in the easy-axis case, and characterize the scaling of
long-lived oscillations around this stationary profile, both in LL and
XXZ.  Section \ref{sec:Delta=1} is devoted to the isotropic models
($\Delta=1$). We compare the $z$-component of the magnetization
profile in both problems, and we confirm the finding of
\cite{misguich_dynamics_2017} and~\cite{gamayun_domain_2019}
concerning respectively a logarithmic correction to diffusion in the
quantum case and in the classical LL case. We also consider the $x$ and $y$
components of the LL magnetization, which show small amplitude
oscillations extending beyond the diffusion scale, up to $r/t$ of
order one. While the in-plane magnetization is zero (by symmetry) in
the quantum problem, the in-plane magnetization of the LL problem can
be quantitatively compared with spin-spin {\em correlations} in the
quantum chain (Sec.~\ref{ssec:correl}). An heuristic argument concerning
the (logarithmic) growth of the entanglement entropy in the spin-1/2
model is also given.  In Sec.~\ref{sec:var_NLS} we exploit the mapping
between the isotropic LL equation and the NLS equation to construct a
variational solution which captures several aspects of the isotropic
domain problem: not only the diffusive-like expansion of the $z$
component of the magnetization profile, but also the larger scale
behavior of several geometrical quantities like the curvature or the
torsion associated to the LL to NLS mapping. Finally, the relevance of
the self-similar solutions of the isotropic LL equation to the
present domain-wall problems is discussed.  The mapping from LL to NLS
is recalled in Appendices~\ref{sec:LL2NLS} and \ref{sec:NLS2LL}.
An explicit calculation of self-similar solution of the LL equation is
given in Appendix~\ref{sec:kummer}.

\section{The model: domain wall problem in the XXZ chain}

\subsection{XXZ model}

We study the evolution of a quantum XXZ spin-1/2 chain, prepared at
time $t=0$ in a domain wall state $\ket{\rm DW}=\ket{
  \uparrow\uparrow\cdots\uparrow\uparrow\downarrow\downarrow\cdots\downarrow\downarrow}$
where all the spins in the left half of the system are ``up''
($S^z=\frac{1}{2}$) and those in the right half are ``down''
($S^z=-\frac{1}{2}$).  At time $t>0$ the wave function then evolves
according to
\begin{equation}
\ket{\psi(t)}=\exp\left(-i\hat Ht\right)\ket{\rm DW} \label{eq:psi(t)},
\end{equation}
where the Heisenberg Hamiltonian $\hat H$ is that of a XXZ chain of
length $L$ with open boundary conditions,
\begin{eqnarray}
  \hat H&=&-\sum_{r} 
\left(\hat S^x_r \hat S^x_{r+1}+\hat S^y_r \hat S^y_{r+1} 
+\Delta \left[\hat S^z_r \hat S^z_{r+1}-\frac{1}{4} \right] \right)
  \label{eq:H}
  \end{eqnarray}
  and $\Delta$ is the anisotropy parameter.\footnote{The constant term
    $-\frac{1}{4}\Delta$ is introduced here to set at zero the energy
    of a fully polarized state in the $z$ direction. The global minus
    sign in the definition of $\hat H$ is here to simplify the connection
    with the classical {\em ferromagnetic} LL description, but it has
    no influence on the dynamics when starting from $\ket{\rm DW}$ at
    $t=0$.} $r\in \mathbb{Z}$ labels the lattice sites, but in the
  next section $r$ will be treated as a continuum variable.

  This problem was first studied by Antal {\it et
    al.}~\cite{antal_transport_1999,antal_logarithmic_2008} in the
  free fermion case ($\Delta=0$), where an exact analytical solution
  for the long-time limit of the magnetization profile was obtained. A
  few years later the problem with $\Delta\ne0$ was studied
  numerically by Gobert {\it et al.} \cite{gobert_real-time_2005}
  using the time-dependent density-matrix renormalization group
  (DMRG).  For $|\Delta|<1$ an exact solution for the long-time limit
  of the magnetization profile was recently
  obtained~\cite{collura_analytic_2018} using
  GHD~\cite{bertini_transport_2016,castro-alvaredo_emergent_2016}.
  Despite some recent analytical progresses~\cite{stephan_return_2017},
  the precise scaling of the front remains unknown at the Heisenberg
  point $\Delta=1$, even though recent large-scale numerics suggest
  that it shows some diffusive behavior, possibly with a
  multiplicative logarithmic correction~\cite{misguich_dynamics_2017}.

  In the following, the data for the XXZ spin chain are obtained using
  a tDMRG
  algorithm~\cite{white_real-time_2004,daley_time-dependent_2004},
  implemented using the iTensor library~\cite{itensor}, as in
  Ref.~\cite{misguich_dynamics_2017}. Technically, we typically use
  systems of size $L=800$ sites, Trotter time-step $dt=0.2$
  (matrix-product operator scheme
  $W^{II}$~\cite{zaletel_time-evolving_2015} at order
  4~\cite{bidzhiev_out-equilibrium_2017}), a matrix-product state
  (MPS) truncation parameter equal to $10^{-10}$ or $10^{-11}$ and
  maximum bond dimension from $1000$ to $2500$.
  These state-of-the art simulations are pushed up to $t=300$.
  On the LL side, the
  calculations are performed using the software Maple (pdsolve) with
  systems size up to $L=3200$, and space and time discretization steps
  $dx=dt$ from $0.01$ to $0.05$. We checked that, at the scale of the plots,
  the results presented here are essentially free of finite-step of finite
  size errors.

\subsection{Anisotropic Landau Lifshitz equation}
\label{ssec:aniLL}

From a different perspective, one can study an analogous domain-wall
problem for a {\em classical} spin model.  To this end, we consider
the anisotropic Landau-Lifshitz (LL) equation. It describes the
(precession) dynamics of a classical XXZ model in the continuum
limit~\cite{kosevich_magnetic_1990,lakshmanan_fascinating_2011}:
\begin{equation}
 \vec \Omega_t = \vec \Omega \wedge \left[ \vec \Omega_{rr} 
+  \delta \left(\Omega^z \vec e_z  \right)\right] ,
 \label{eq:LL}
\end{equation}
where $\delta$ is the anisotropy parameter, and $r$ has become a
continuous variable.\footnote{We use	 here the compact notations
  $\partial_r^2 (\Omega^\alpha)\to \Omega^\alpha_{rr}$ and $\partial_t
  \Omega^\alpha\to \Omega^\alpha_t$ for the space and time
  derivatives.}  Note that the $z$ component of the equation above can
be re-written
\begin{equation}\label{eq:Ia}
 \Omega^z_t+I_r = 0,
\end{equation}
where
\begin{equation}
I(r,t) =\Omega^y \Omega^x_{r}-\Omega^x \Omega^y_{r}, \label{eq:Ib}
\end{equation}
is the current associated to the $z$-component of the magnetization.

Since we wish to make some {\em quantitative} comparison between the
local magnetization $\langle S^z \rangle$ of the XXZ chain with
$\Omega^z$, we set the norm of the LL magnetization vector to $|\vec
\Omega|=1/2$, and one can easily check that the anisotropy should be
set to
\begin{equation}\label{eq:delta}
\delta=2(\Delta-1).
\end{equation}
The above equations of motion can also be obtained from the following
classical Hamiltonian:
\begin{eqnarray}
H_{\rm LL}&=&\frac{1}{2}\int \left[
\left(\partial_r \vec \Omega\right)^2
- \delta \left\{(\Omega^z)^2-\frac{1}{4}\right\}
\right]dr \label{eq:LL-H}.
\end{eqnarray}
The constant term ensures, as in the quantum case, that a uniform
state polarized in the $z$ direction has zero energy. $\delta>0$
corresponds to easy-axis cases, $\delta<0$ to easy-plane, and
$\delta=0$ to the isotropic model. The isotropic LL model is also
often called the classical Heisenberg model.

As proposed in Refs. \cite{gamayun_domain_2019,mallick_2017}, for the
LL model a natural counter part of the lattice domain wall state is
the following smooth initial condition:
\begin{equation}
\label{eq:init_cond}
\vec M(r,t=0)=2\, \vec\Omega(r,t=0)=
\left(
\begin{matrix}
1/\cosh(ar) \\
0 \\
-\tanh(ar)
\end{matrix}
\right).
\end{equation}
It describes some smooth domain wall which interpolates between $\vec
\Omega = \frac{1}{2} \vec e_z$ at $r\to-\infty$ and $\vec \Omega =
-\frac{1}{2} \vec e_z$ at $r\to+\infty$; $a$ is the inverse width of
the initial profile.

\begin{figure}[h]\center\includegraphics[height=0.35\linewidth]{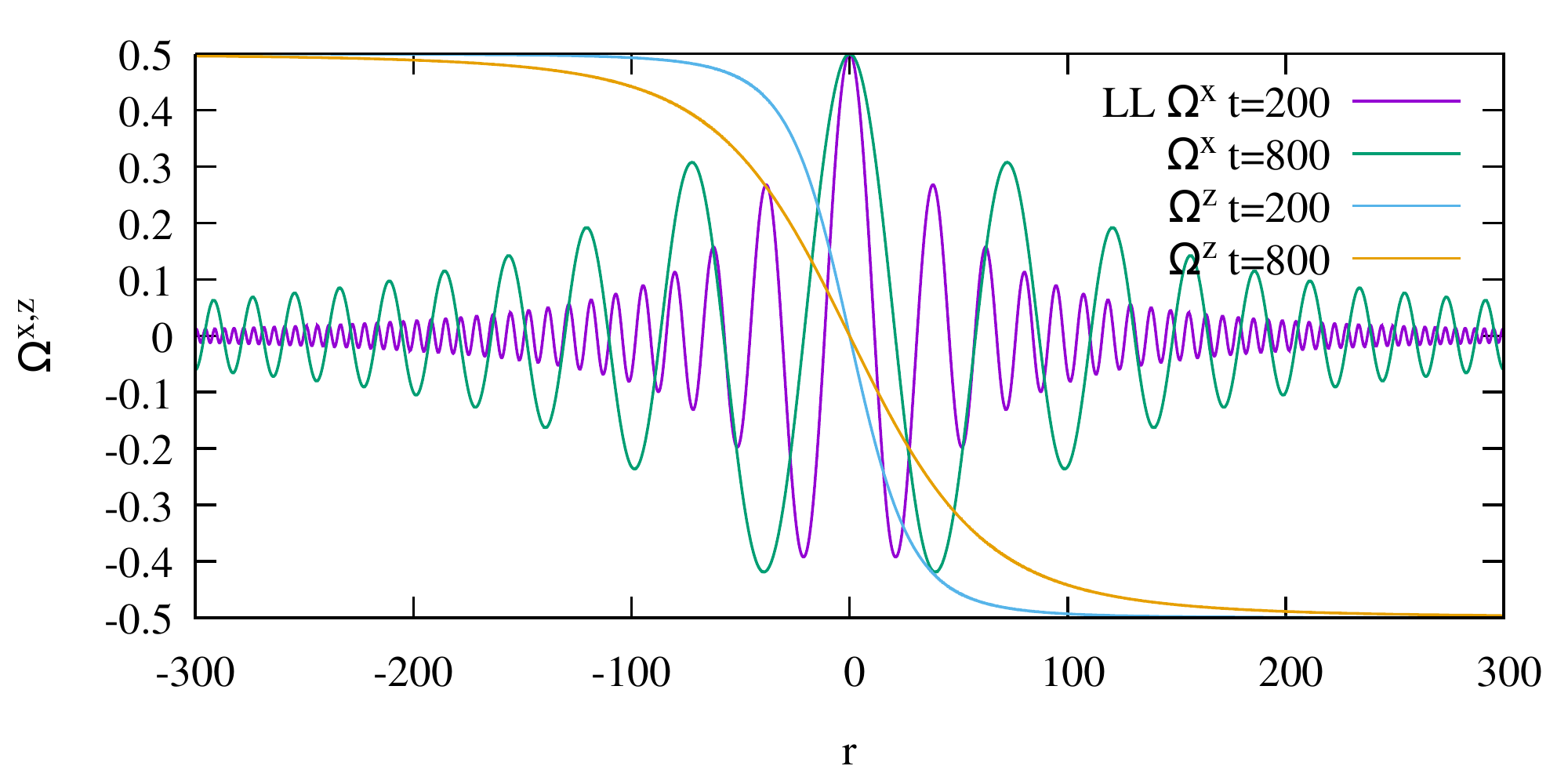}
  \caption{$\Omega^x$ and $\Omega^z$ magnetization profiles for the
    isotropic LL model ($\Delta=1$) at $t=200$ and $t=800$ (initial
    condition given in Eq.~\eqref{eq:init_cond}).}
\label{fig:S1S3D1}
\end{figure}

For $\delta\ne 0$ it is convenient to define a new space variable
$\zeta=r\,|\delta|^{\frac{1}{2}}$ and a new time variable
$\tau=\frac{1}{2}|\delta|\, t$, so that the LL equation becomes:
\begin{equation}
\vec{M}_{\tau}=\vec{M}\wedge\vec{H}_{\rm eff}\; ,
\quad\mbox{where}\quad
\vec{H}_{\rm eff}=
\vec{M}_{\zeta\zeta}+\mathrm{sign}(\delta)\, M_z\, \vec{e}_z.
 \label{eq:rescaledLL}
\end{equation}
Now, the inverse width $a$ of the initial conditions is changed to
$\tilde a=a/\sqrt{|\delta|}$ in the variable $\zeta$. In other words,
in the case $|\delta|\ll 1$ we are interested in, we may consider
Eq.~\eqref{eq:rescaledLL} with an anisotropy parameter fixed to
$\pm1$, but with a very narrow initial profile in terms of $\zeta$
($\tilde a \to \infty$).

\section{Energy conservation and classical behavior close to $\Delta=1$}
\label{sec:energy}

We give here a heuristic argument based on energy conservation,
arguing that the behavior of the quantum XXZ chain for $\Delta$ close
to 1 gets closer and closer to that of a classical LL problem as the
energy and the $\langle S^z \rangle$ profile spread over larger
distances.

\subsection{Energy in the domain wall problem}
\label{ssec:energy_cons}

For the quantum XXZ chain [Eq.~\eqref{eq:H}], the energy density on
one bond can be written as
\begin{equation}
 \hat H_r=H^{\rm iso}_r  +(\Delta-1) V_r,
\end{equation}
with
\begin{eqnarray}
H^{\rm iso}_r&=&\frac{1}{4}-\vec S_r \cdot \vec S_{r+1} \nonumber \\
V_r&=&\frac{1}{4}-S_r^z S_{r+1}^z.\nonumber 
\end{eqnarray}
The largest eigenvalue of $\vec S_r \cdot \vec S_{r+1}$ is
$\frac{1}{4}$, corresponding to the triplet states (three-fold
degenerate).  As for $S^z_r \hat S^z_{r+1}$, its largest eigenvalue is
also $\frac{1}{4}$, and it corresponds to $|\!\uparrow\uparrow\rangle$
and $|\!\downarrow \downarrow\rangle$ (two-fold degenerate).  So, the
mean values of two terms $H^{\rm iso}_r$
and $V_r$ appearing in $\hat H_r$
are necessarily $\geq 0$.  The total mean energy $E_{\rm XXZ}=\bra{\psi(t)}\hat
H \ket{\psi(t)}$ is conserved along the time evolution and its value
is easily computed at $t=0$, where only the central bond contributes:
$E_{\rm XXZ}=\bra{\rm DW}\hat H \ket{\rm DW}=\frac{1}{2}\Delta$.

We now assume that there exists some length scale $R_\Delta(t)$ over
which the energy is spread after time $t$.  We further make the
simplifying assumption that the energy density is finite and that it
varies smoothly with position in this region, and that the energy
outside $\left[-R_\Delta(t),R_\Delta(t) \right]$ is negligible.  The total energy
$E_{\rm XXZ}$ can then be written as $E_{\rm XXZ}=2e(t)R_\Delta(t)$
where
\begin{eqnarray}
e(t)&=&\left< H^{\rm iso} \right> + (\Delta-1) \left< V \right>
\label{eq:e}
\end{eqnarray}
is a mean energy density, and $\langle \cdots \rangle$ denotes a
quantum average as well as a spatial average over the bonds in the
nontrivial region of the system, where the energy is distributed.  We
conclude that
\begin{equation}
\frac{\Delta}{4R_\Delta(t)}=\left<H^{\rm iso} \right> + \frac{\delta}{2}\; \left< V \right>.
\label{eq:energy_cons}
\end{equation}
Let us examine the implications of this relation when $R_\Delta(t)$
tends to infinity.
\begin{itemize}
\item If $\Delta=1$ ($\delta=0$), we
  know~\cite{misguich_dynamics_2017} that the $\langle S^z\rangle$
  profile extends in a diffusive way, with logarithmic corrections.
  This implies that the energy spreads over a distance which is at
  least as large as $\mathcal{O}(\sqrt{t})$, and thus
  $R_\Delta(t)\to\infty$. The energy density therefore tends to zero
  everywhere when $t\to \infty$.  We will see later (in
  Sec.~\ref{ssec:energy_density}) that the energy density in the
  isotropic spin chain goes to zero as $1/t$,
  possibly with logarithmic corrections.  From
  Eq.~\eqref{eq:energy_cons} we get that $\langle H^{\rm iso} \rangle$ vanishes when $t\to\infty$. We conclude that,
  in the nontrivial region $[-R_\Delta(t),R_\Delta(t)]$, the
  nearest-neighbor correlations asymptotically become that of a
  ferromagnetic state. The spins thus become locally aligned, but the
  direction of the magnetization is unconstrained.
\item If $\Delta>1$ ($\delta>0$) the two terms in the right-hand side
  of Eq.~\eqref{eq:energy_cons} are positive. If we further assume
  that $R_{\Delta}(t)$ is large, then $\langle H^{\rm iso}\rangle$ (as well as $\delta\langle V\rangle$) will be small, of order
  $\mathcal{O}\left(R_{\Delta}(t)^{-1}\right)$. We will see below
  (Sec.~\ref{sec:easy_axis}) that 
  the $\langle S^z\rangle$  profile extends over a distance
  of the order of $\delta^{-\frac{1}{2}}$. So, even though 
  $R_\Delta(t)$ may not 
  diverge when $t\to\infty$ at fixed $\Delta>1$, we have
  $\lim_{t\to\infty} R_\Delta(t)  > \mathcal{O}\left(\delta^{-\frac{1}{2}}\right)$.  So, in
  the limit of a weak easy-axis anisotropy ($\delta\ll1$), the state
  again approaches locally a ferromagnetic state.
\item If $\Delta<1$ we will see that there is a ballistic propagation
  of the front at velocity $v_{\rm XXZ}=\sqrt{1-\Delta^2}$, and thus
  $R_\Delta(t) \simeq t \;v_{\rm XXZ} \to \infty$. So, at long times
  $\langle H^{\rm iso}\rangle\simeq (1-\Delta)
  \langle V\rangle$ and thus $0 \leq
  \langle H^{\rm iso}\rangle \leq
  \frac{1}{4}|\delta|$ since $\langle V \rangle \leq
  \frac{1}{2}$.  We see here that when $\Delta$ approaches $1^-$, the
  nearest-neighbor correlations again become that of a ferromagnetic
  state.
\end{itemize}

In all the cases above, where the nearest-neighbor correlation become
asymptotically that of a ferromagnetic state, we expect that the spins
will become ferromagnetically aligned over {\em large
  distances}. These large, almost ferromagnetic, segments of the chain
will have a large total spin and should therefore behave
semi-classically.  This should occur at large times for $\Delta=1$,
but also if we consider the regime where one simultaneously takes the
limit of large time and $\Delta\to 1$.  In such situations we
conjecture that the quantum effects as well as the lattice effects
will become weaker and weaker, and that the {\em classical
  (anisotropic) LL description should become quantitatively accurate}
when compared to the XXZ problem. This conjecture is supported by the
data presented below, where we compare the numerical solutions of the
(classical) LL equations to numerical results obtained for the
(quantum) XXZ spin chain with $\Delta$ close to 1.

Of course the short-time dynamics of the XXZ model is not classical,
and we should therefore expect a quantitative agreement between the
lattice quantum problem and the classical LL model only for quantities
that are independent of the short-distance properties of the initial
domain wall.  Another important consequence of the lattice is the
existence of a maximum Lieb-Robinson velocity in the quantum chain,
whereas arbitrarily high velocities can in principle be observed in
the continuum limit.
Weak quantum fluctuations will also be present at any finite time,
since the energy density is never strictly zero for
$t<\infty$. Analyzing in detail the quantum corrections to the
classical dynamics is beyond the scope of this paper, but the
numerical data presented in this study show that many observables in
the XXZ chain behave almost classically at long times.

\subsection{Initial width of the LL domain wall}
\label{ssec:a_Delta}

Using Eq.~\eqref{eq:LL-H} one finds that the energy of the initial
condition given in Eq.~\eqref{eq:init_cond} is
\begin{equation}
E_{\rm LL}= \frac{1}{4}(a+\delta/a). 
\label{eq:E}
\end{equation}
Being conserved by dynamics, the value of this energy will constrain
the evolution of the system. In order to be able to make some
quantitative comparison between the dynamics of the quantum system and
that of the classical system, we chose $a$ such that the classical
energy is the same as that of the quantum problem, which is equal to
$E_{\rm XXZ}=\frac{1}{2}\Delta$. In what follows we therefore take
$a(\Delta)$ to be solution of $a+2(\Delta-1)/a=2\Delta$.  This implies
in particular that for $\Delta\to1$ we have $a\to2$.

We will successively discuss the three regimes: easy-plane ($0\leq
\Delta<1$), easy-axis ($\Delta>1$) and finally the isotropic model
($\Delta=1$).

\section{Easy plane regime $\Delta<1$}
\label{sec:easy-plane}

It has recently been shown~\cite{gamayun_domain_2019} that in the long
time limit, the easy-plane LL equation admits solutions of the form
$\Omega^z(r,t)=-\frac{r}{2t}|\delta|^{-\frac{1}{2}}$ for $|r|\leq t
\sqrt{-\delta}$, and $\Omega^z=\pm 1/2$ outside this interval.  Or,
equivalently: $M^z(\zeta,\tau)=-\frac{\zeta}{2\tau}$ for
$|\zeta/\tau|\leq2$.  We give below a simple derivation of this solution.  In Sec.~\ref{sec:vLL_vXXZ_GHD}
we show that, for $\Delta\to1^-$, the above linear profile in the LL
model exactly matches that derived from GHD for the easy-plane XXZ
chain. This is the first manifestation of the asymptotic classical
behavior conjectured in Sec.~\ref{ssec:energy_cons}.

\subsection{Solution of LL -- neglecting dispersion effects}
\label{sec:linear_profile}

When $\vec M$ is parametrized using spherical coordinates
\begin{equation}\label{polarization}
  \vec{M}=
\begin{pmatrix}
\sin\theta\cos\varphi\\
\sin\theta\sin\varphi\\
\cos\theta
\end{pmatrix}\;,
\end{equation}
the Eq.~\eqref{eq:rescaledLL} becomes~\cite{lakshmanan_1976}
\begin{subequations}\label{eq7}
\begin{align}
& \theta_\tau=-2\,\theta_\zeta\, \varphi_\zeta\cos\theta
- \varphi_{\zeta\zeta}\sin\theta\; ,\label{eq7a} \\
& \varphi_\tau=- \cos\theta\left[\varphi_\zeta^2+{\rm sign}(\delta)\right]
+\frac{\theta_{\zeta\zeta}}{\sin\theta}.\label{eq7b}
\end{align}
\end{subequations}
where ${\rm sign}(\delta)=-1$ in the easy-plane regime, ${\rm
  sign}(\delta)=1$ in the easy-axis regime and ${\rm sign}(\delta)=0$
for the isotropic Heisenberg case.

\begin{figure}[h]\center
\includegraphics[height=0.35\linewidth]{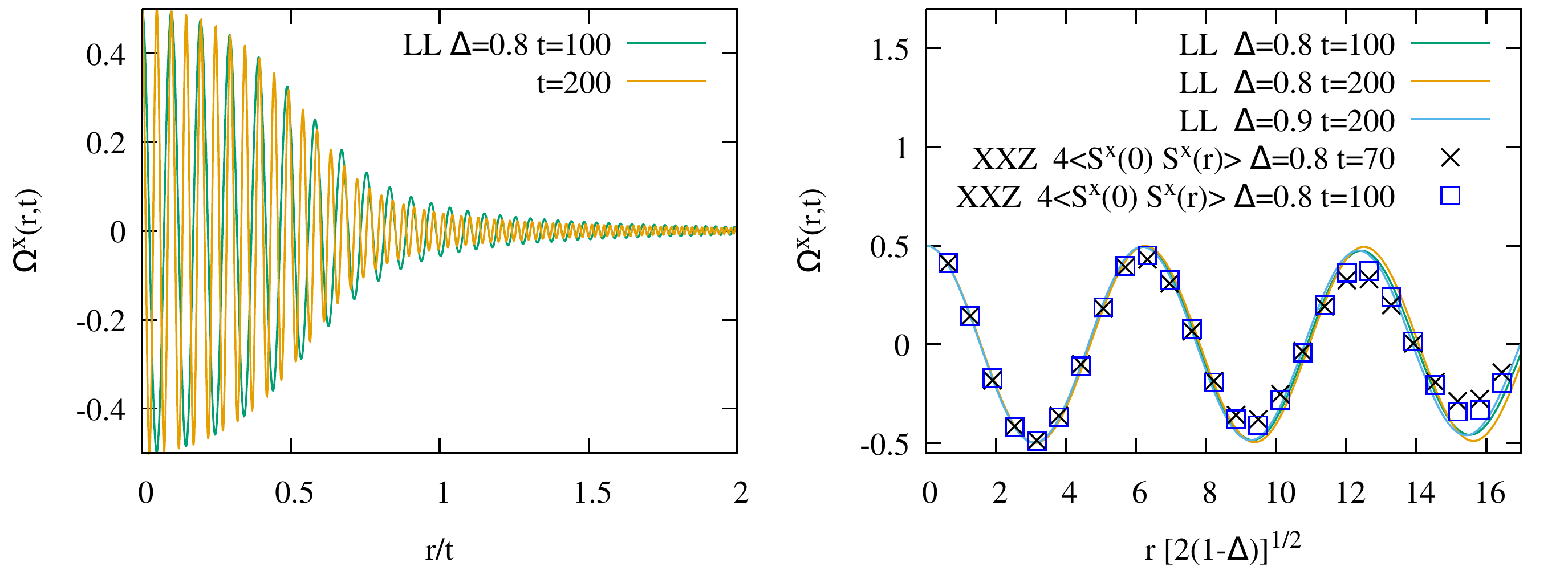}
\caption{ Left panel: $\Omega^x(r,t)$ as a function of the rescaled
  position $r/t$ for $\Delta=0.8$ for different times $t=100$ and
  $t=200$.  The envelope of the oscillations is approximately the same
  at $t=100$ and $t=200$ (this follows from the fact that $\Omega^z$
  scales as $r/t$, see Fig.~\ref{fig:SzDelta_lt1}).  Right panel: when
  plotted as a function of
  $r\sqrt{|\delta|}=r\sqrt{2(1-\Delta)}=\zeta$, the period of the
  first oscillations turns out to be almost independent of time and
  $\Delta$, and is numerically close to $2\pi$. This can be understood
  using the asymptotic solution discussed in
  Sec.~\ref{sec:linear_profile}, which predicts that $|\varphi_\zeta|=1$.
  The crosses are spin-spin correlations in the XXZ chain at
  $\Delta=0.8$, $t=70$ and $t=100$. See Sec.~\ref{ssec:entanglement}
  for the relation between spin-spin correlations in the quantum chain
  and the in-plane component of the LL magnetization.
}\label{fig:SxDelta_lt1}
\end{figure}

The term $\frac{\theta_{\zeta\zeta}}{\sin\theta}$ in \eqref{eq7b}
represents dispersive effects.  We refer, for instance, to the
discussion in \cite{congy_dispersive_2016,ivanov_solution_2017}, where
the so-called ``Riemann problem'' for the easy-plane LL equation was
studied in the context of two-component Bose-Einstein condensates. We
now make the assumption that it can be neglected in the long time
limit.  Next, in view of the $\varphi_\zeta^2-1$ factor, we look for a
solution where $\varphi_\zeta=-1$.  This means that the projection of
$\vec \Omega$ in the $x-y$ plane forms a spiral with a constant pitch,
which is perfectly consistent with the data shown in
the right panel of Fig.~\ref{fig:SxDelta_lt1}.

\begin{figure}[h]\center\includegraphics[width=\linewidth]{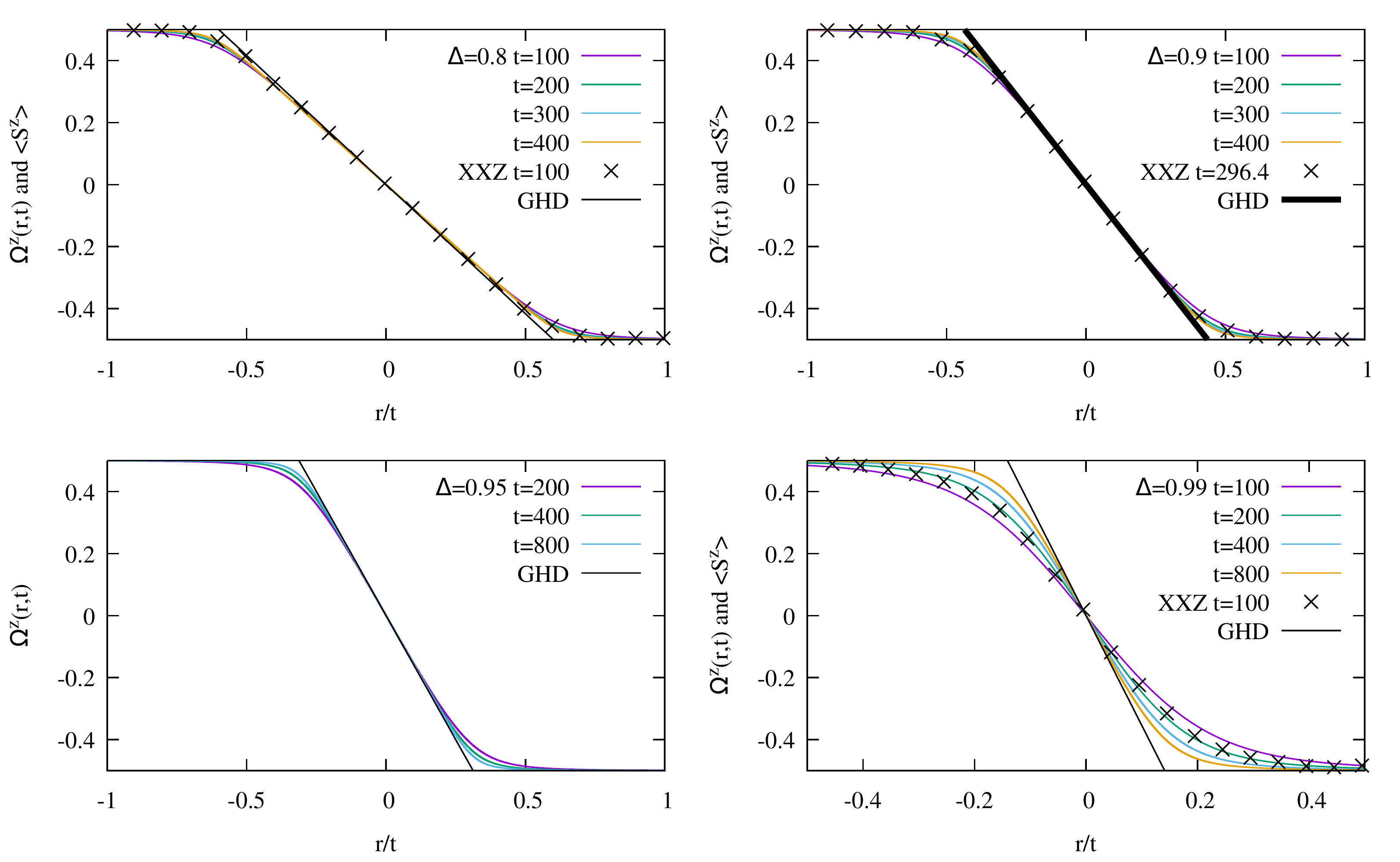}
 \caption{$\Omega^z(r,t)$ from the solution of the classical LL equation,  
as a function of the rescaled position $r/t$ for $\Delta=0.8$, 
$\Delta=0.9$, $\Delta=0.95$ and $\Delta=0.99$
 and  for different times from $t=100$ to $t=800$.
 The black lines, labeled GHD, represent the long-time limit of the profile 
for the XXZ chain, $\langle S^z\rangle=-\frac{r}{2t}
\left(1-\Delta^2\right)^{-1/2}=
-\frac{r}{2t}(v_{\rm XXZ})^{-1}$~\cite{collura_analytic_2018},
 as predicted by GHD.
 In the panels associated to $\Delta=0.8,9$ and $\Delta=0.99$, the black crosses represent tDMRG data for the XXZ chain.
}\label{fig:SzDelta_lt1}
\end{figure}

Then Eq. \eqref{eq7b} simplifies to $\varphi_\tau=0$, and \eqref{eq7a} becomes
\begin{equation}
 \theta_\tau=2\,\theta_\zeta\,\cos\theta.
\end{equation}
For $|\zeta/\tau|\leq2$ an obvious solution of the equation above is then
\begin{equation}
 \cos\theta=-\frac{\zeta}{2\tau},
\end{equation}
which corresponds to $M^z=-\frac{\zeta}{2\tau}$. One can check a
posteriori that, in this solution, $\theta_{\zeta\zeta}$ is of order
$\mathcal{O}(\tau^{-2})$ and thus becomes negligible at long times.
For $|\zeta/\tau|\leq2$, the solution lies at the boundary of
the hyperbolicity domain of the easy axis LL system, and is sometimes
referred to as a ``vacuum region'' \cite{Gurevich_Krylov_1987}. Indeed,
for this range of values of $\theta$ and $\phi_\zeta$, the model can
be mapped onto a system of shallow-water Kaup-Boussinesq equations
\cite{ivanov_solution_2017} for which the condition $|\varphi_\zeta|=1$
corresponds to the absence of the fluid \cite{congy_KB_2017}. For
$|\zeta/\tau|>2$ the linear profile makes room for a constant
magnetization with $\cos\theta=\pm1$.
Going back to the original $r$ and $t$ variables, this corresponds to
a front which propagates at the velocity
\begin{equation}
v_{\rm LL}=\sqrt{-\delta}=\sqrt{2(1-\Delta)}.
\end{equation}

As can be seen in
Fig.~\ref{fig:SzDelta_lt1},
the transient time needed to converge to
such a profile increases when $\Delta$ approaches $1^-$.
Indeed, at $\Delta=0.99$ and $t=100$ for instance, the LL profile
(as well as the XXZ one) is still
relatively far from the asymptotic result.
These finite
time corrections are the largest in the region corresponding to
the $r/t\simeq v_{\rm LL}$ where the linear part of the profile
connects to the constant part.
Looking at $\Omega^x$
(Fig.~\ref{fig:SxDelta_lt1}), the numerical solutions also confirm
that $\varphi_\zeta\simeq -1$, as predicted above.

\subsection{Easy plane LL and GHD}
\label{sec:vLL_vXXZ_GHD}

For the XXZ spin chain the exact form of the $\langle S^z\rangle$
profile has been computed in the framework of GHD.  For a generic
values of $\Delta$~\footnote{$\Delta$ may be parametrized as
  $\Delta=\cos(\Theta)$. We then call {\em generic} the values of
  $\Delta$ for which the angle $\Theta$ is {\em not} of the form
  $\Theta=\pi Q/P$, with $Q$ and $P$ coprime integers satisfying
  $1\leq Q<P$.}  it turns out that this profile is a simple linear
function of $r/t$~\cite{collura_analytic_2018}:
\begin{equation}
 \langle S^z\rangle=-\frac{r}{2 v_{\rm XXZ} t}.
\end{equation}
with a velocity given by:
\begin{equation}
 v_{\rm XXZ} =\sqrt{1-\Delta^2}.
\end{equation}
So, the LL and XXZ problems both show a linear profile for the $z$
component of the magnetization.

But, in addition, we observe that the slopes of the profiles become
{\em identical} in the limit $\Delta\to1$, since $v_{\rm XXZ}/v_{\rm
  LL}=\sqrt{(\Delta+1)/2}$.  We argue that this is not an accidental coincidence but
a manifestation of the asymptotic classical behavior discussed in
Sec.~\ref{ssec:energy_cons}.  Naturally, this match relies on the fact
that, on the LL side, the initial configuration has the same energy as
the domain wall for the quantum chain.

While the emergence of a linear profile is quite easily understood for
the LL model (Sec.~\ref{sec:linear_profile}), the fact that such a
simple profile emerges from such a complicated model as the XXZ chain
(for a generic value of $\Delta$) and from the the GHD equations is
quite remarkable. Clearly, the asymptotic equivalence with LL when
$\Delta\to1^-$ -- as argued in Sec.~\ref{sec:energy} -- sheds some
light on this question.
It would also be interesting to compare further the finite-time behavior of the LL magnetization profile
with that of the XXZ chain, in the light
of the finite-time corrections to the asymptotic profile close to the edge 
at $r/t=v_{\rm XXZ}$, as studied in Refs.~\cite{collura_analytic_2018,stephan_free_2019}.

\section{Perturbative expansion}
\label{sec:sw}

We consider here the classical LL problem and focus on the region
where $\vec M$ is close to $-\vec e_z$.  At a given time, this is
supposed to hold for sufficiently large $r$.  We can therefore write
\begin{equation}
  \vec M=2\, \vec \Omega(r,t)=\alpha(r,t) \vec e_x + 
\beta(r,t) \vec e_y - \sqrt{1-\alpha^2-\beta^2}\vec e_z
\end{equation}
with $\alpha$ and $\beta \ll1$. The equation \eqref{eq:LL} can be
expanded to first order in $\alpha$ and $\beta$.
Using the complex variable $z=\alpha+i\beta$, the linearized equation reads
\begin{equation}\label{eq:z_t}
2z_t=-iz_{rr}+i\delta z, 
\end{equation}
and admits the following plane wave solutions
\begin{eqnarray}\label{eq:zsw}
z(r,t)&=&z_0 \exp\left(i\left[\omega(k) t -kr\right] \right),
\end{eqnarray}
where the frequency $\omega(k)$ is related to the wave vector $k$
through the dispersion relation
\begin{equation}
 \omega(k) = \frac{1}{2} \left( k^2 + \delta \right).
\end{equation}
These are right moving solutions, and the left moving ones can be
obtained from the expressions above by replacing $k$ by $-k$ in the
expression of $z$ [Eq.~\eqref{eq:zsw}].

Eq.~\eqref{eq:z_t} being linear, a general solution can be constructed
by superposition of these plane waves:
\begin{equation}
z(r,t)=\int_{-\infty}^{\infty} \frac{dk}{2\pi} z_0(k) 
\exp\left(i\left[\omega(k) t -kr\right] \right)
\end{equation}
where $z_0(k)$ is the Fourier transform of the initial condition at
$t=0$.  Consider now the long-time and long-distance limit, with
$x=r/t$ fixed:
\begin{equation}
z(r,t)=\int_{-\infty}^{\infty} \frac{dk}{2\pi} z_0(k) 
\exp\left(it\left[\omega(k) -kx\right] \right)
\end{equation}
When $t\to\infty$ the integral above is dominated by
the vicinity of a saddle point located at a wave vector $k_0$ determined by
\begin{eqnarray}
  \omega'(k_0)&=&x,
\end{eqnarray}
which gives
\begin{equation}
    k_0=x=r/t.
\end{equation}
The saddle point integration then gives
\begin{equation}
z(r,t)\simeq \sqrt{\frac{2i\pi}{t}}\frac{z_0(k_0 )}{2\pi}  \exp\left[-ir^2/(2 t)+it\delta/2\right].
\label{eq:alpha_r_t}
\end{equation}
It is interesting to note that the complex phase in the equation above
is independent of the initial condition; it describes (small)
oscillations around the $z$ direction with an angle (in the $x$-$y$
lane) varying as $\varphi(r,t)=-r^2/(2t)+t\delta/2$. This is in agreement with the
data plotted in Fig.~\ref{fig:S1_r2} (bottom panel) for $\Delta=1$
($\delta=0$) or Fig.~\ref{fig:SxDelta_gt1} at $\Delta=1.2$.

\begin{figure}[h]\center
\begin{picture}(15,8.5)
\put(0,0){\includegraphics[width=\linewidth]{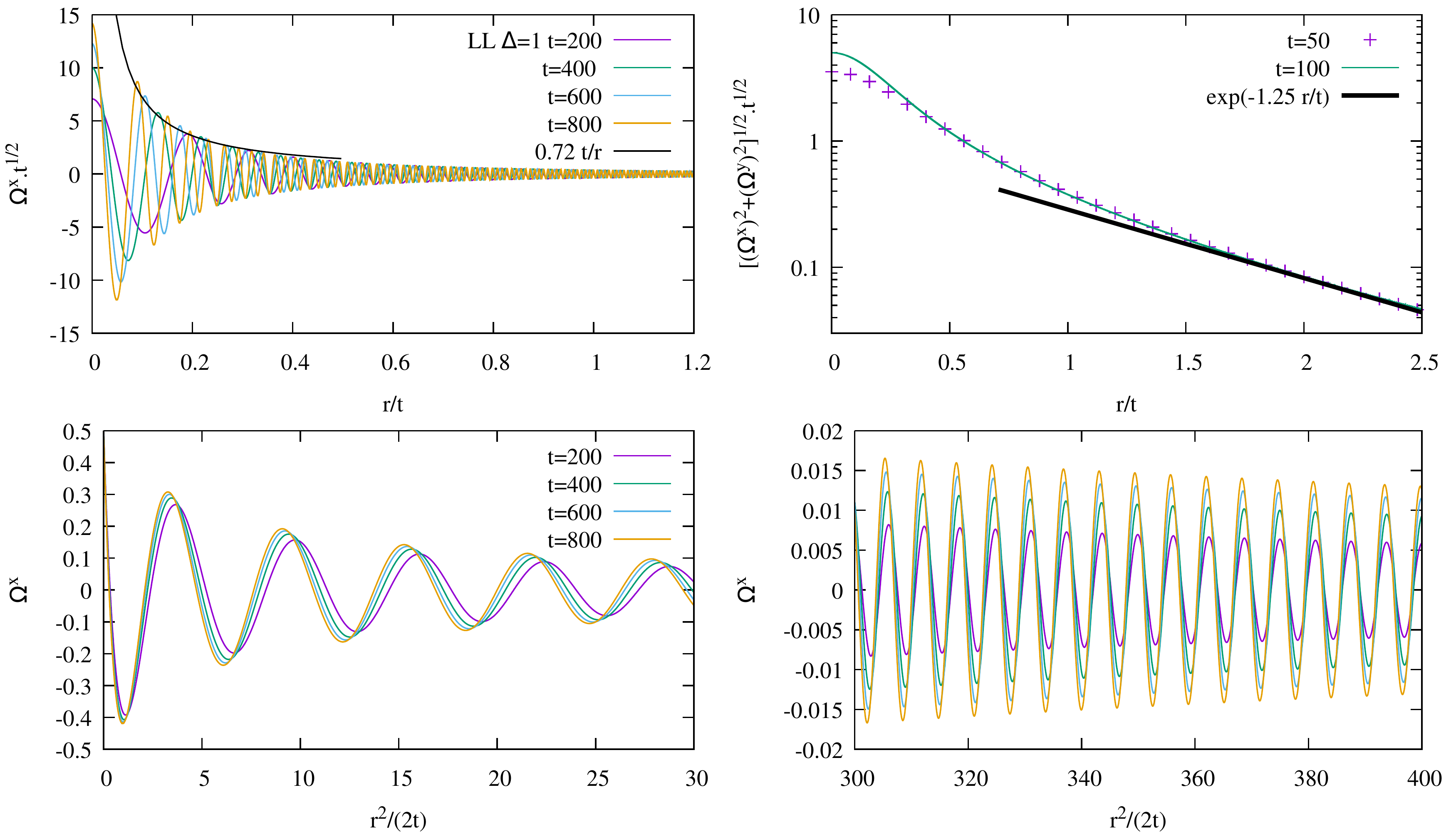}}
\put(6.3,5.4){\large(1)}
\put(9,5.4){\large(2)}
\put(6.3,1.1){\large(3)}
\put(13.8,1.0){\large(4)}
\end{picture}
\caption{ Panel 1 : $\Omega^x \sqrt{t}$ for $\Delta=1$ as a function
  of $r/t$ (LL). With this scaling the different curves ($t=200$, 400,
  600 and 800) appear to have the same envelope. The thin black line
  is a fit of the envelope in the region where $r$ is of the order of
  $\sqrt{t}$ (thus small $r/t$). In this region, the envelope of
  $\Omega^x \sqrt{t}$ appears to scale as $\sim t/r$.  Panel 2: $\sqrt{
    (\Omega^x)^2+(\Omega^y)^2} \sqrt{t}$ in log scale.  The thick black
  line is a fit in the region of large $r/t$, where the amplitude of
  the magnetization in the $x-y$ plane decays exponentially.  This plot
  suggests that $\sqrt{
    (\Omega^x)^2+(\Omega^y)^2}=\sqrt{1/4-(\Omega^z)^2}$ scales as
  $\exp(-cr/t)/\sqrt{t}$ for large $r/t$, which is consistent with
  Fig.~\ref{fig:S3logD1}.  Panels 3 and 4: $\Omega^x$ as a function of
  $r^2/(2t)$. With this scaling, the period of the oscillations is almost the
  same for the four curves, and is found to be numerically close to
  $2\pi$.  This behavior is observed both for small $r^2/t$ (panel 3),
  as well as large $r^2/t$ (panel 4).  $\Omega^x$ can thus be
  described by an oscillatory factor $\cos(r^2/(2t)+{\rm cst})$, in
  agreement with Eq.~\eqref{eq:alpha_r_t}.  }\label{fig:S1_r2}
\end{figure}
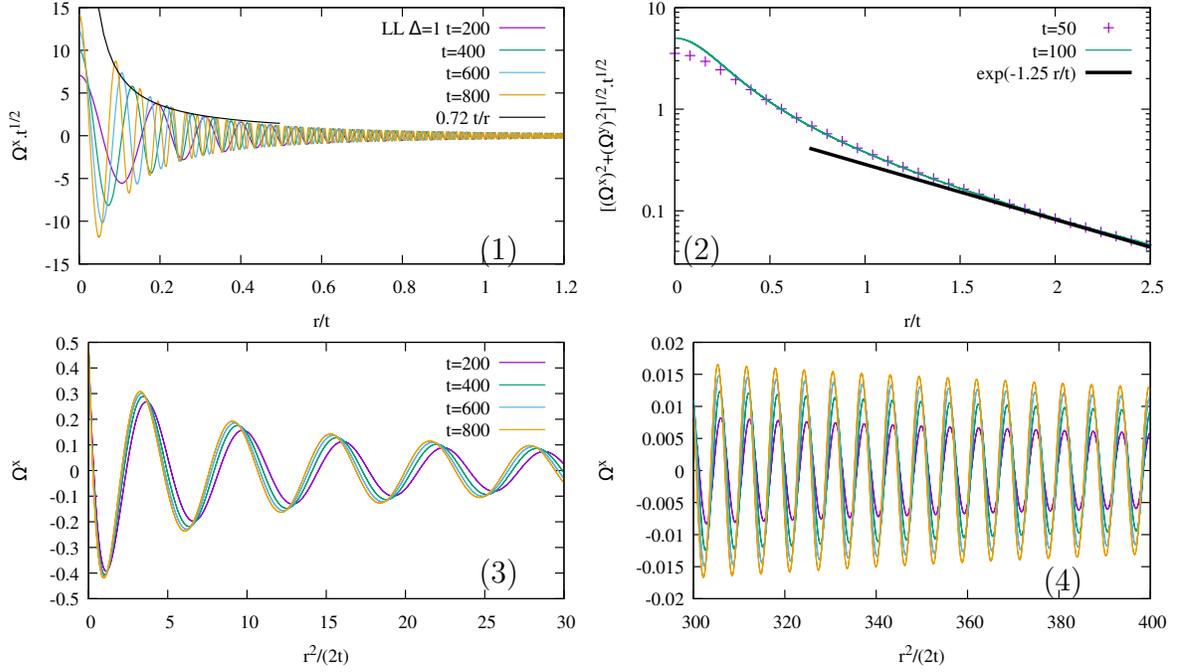
\begin{figure}[h]\center
 \includegraphics[height=0.35\linewidth]{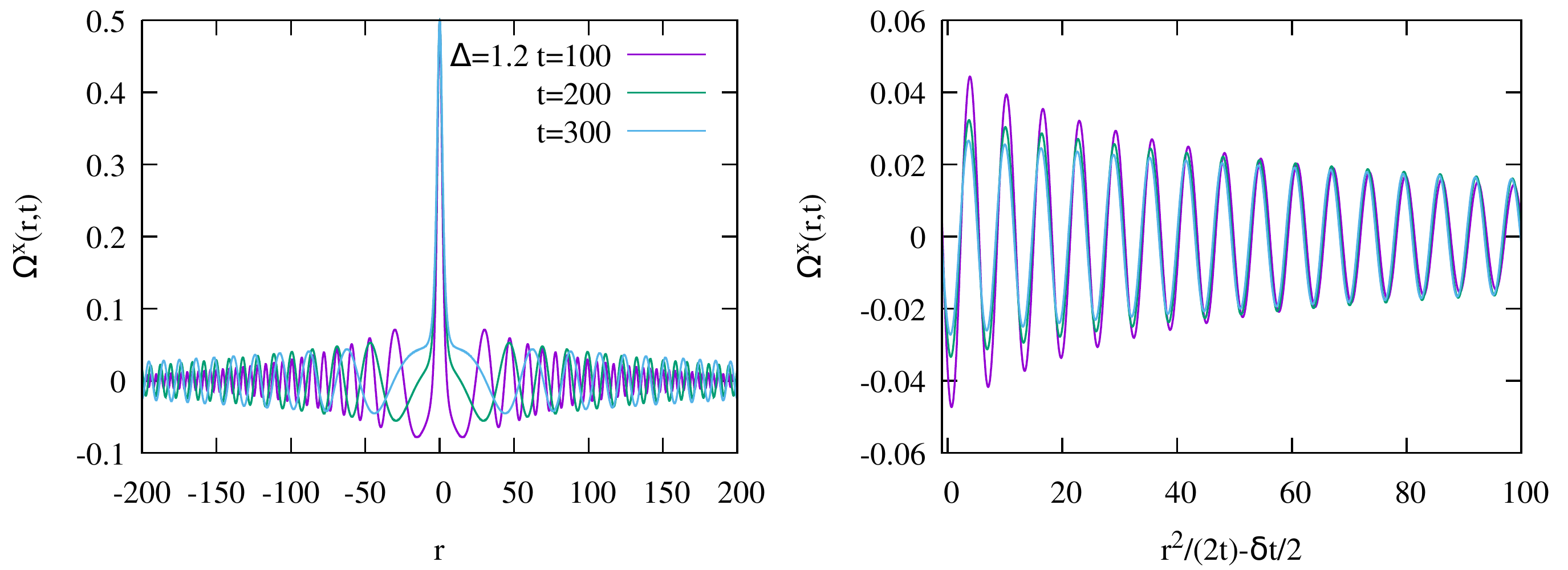}
 \caption{ Large distance behavior of $\Omega^x(r,t)$ for $\Delta>1$ and three different times.
   Left panel: $\Omega^x(r,t)$ as a function of the position $r$ for
   $\Delta=1.2$. Right: Same data plotted as a function of $r^2/(2t) -
   t\delta/2$ with $\delta$ given in Eq~\eqref{eq:delta}. The latter shows that the phase of the oscillations is
   well described by the argument of the exponential in
   Eq.~\eqref{eq:alpha_r_t}.  }\label{fig:SxDelta_gt1}
\end{figure}

The initial condition given by Eq.~\eqref{eq:init_cond} does not
correspond to a small $|z|$ close to the origin at $r=0$, but let us
nevertheless consider the dynamics obtained with the linear equation
[Eq.~\eqref{eq:z_t}], starting from this initial condition.  The
Fourier transform $z_0(k)$ of the initial condition can be computed
explicitly, and also takes the form of the inverse of an hyperbolic
cosine:
\begin{equation}
 z_0(k)=\frac{1}{2}\int_{-\infty}^{\infty} \frac{dr}{\cosh(ar)}\exp\left(ikr\right)
=\frac{\pi}{2a}\frac{1}{\cosh(\pi k/(2a))}.
\end{equation}
plugging the equation above into Eq.~\eqref{eq:alpha_r_t} gives
\begin{eqnarray}
z(r,t)&\simeq& \sqrt{\frac{2i\pi}{t}}\frac{1}{4a\cosh(\pi r/(2at))}
\exp\left[-ir^2/(2 t)+it\delta/2\right].
\label{eq:alpha_cosh}
\end{eqnarray}
In this linear approximation, the amplitude $|z|$ appears to be a
function of $r/t$.  This is in agreement with the upper panels of
Fig.~\ref{fig:S1_r2}, which show that the envelope of the oscillations
in $\Omega^x\times \sqrt{t}$ is a function or $r/t$.  The analytical
result shows that this amplitude scales as $|z|\sim \exp\left(-\pi
  r/[2at]\right)/\sqrt{t}$ for large $r/t$, which is also in agreement
with the data (upper right panel of Fig.~\ref{fig:S1_r2}).  The
numerical value of the constant $c\simeq 1.25$ obtained by a fit is
however different from the constant $c'=\pi /(2a)=\pi/4$ predicted by
the analytical argument above.  This quantitative discrepancy for the
rate of the exponential decay is presumably due to the fact that the
initial condition is not in the linear regime close to the origin.

\begin{figure}[h]\center
 \includegraphics[height=0.35\linewidth]{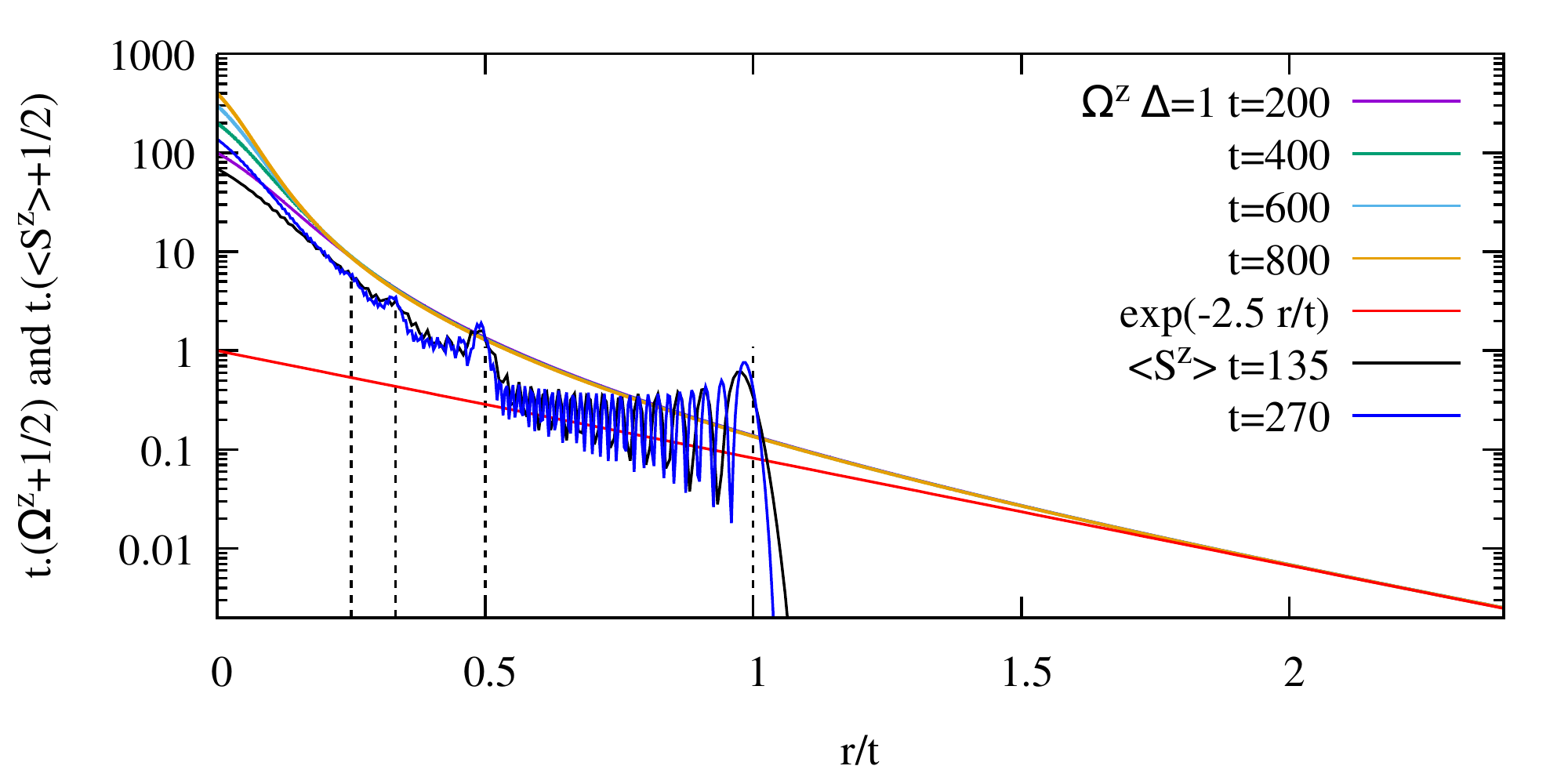}
 \caption{Zoom on the tail of the profile for $\Delta=1$. For the
 LL model, we plot the quantity $t\cdot(\Omega^z+1/2)$.
 The almost perfect collapse of the curves for different times shows
 that, at long times and large distances, we have $\Omega^z+1/2 \simeq
 \frac{1}{t}\exp(-cr/t)$.  For XXZ the black and blue lines correspond to
 the quantity $\left<S^z_r\right>+\frac{1}{2}$ in the Heisenberg
 chain (tDMRG) at $t=135$ and $t=270$ (the two curves are almost on top
 for of each other). The data are consistent with
 the scaling observed for LL. The vertical dotted lines are at
 $r/t=\frac{1}{4}, \frac{1}{3}, \frac{1}{2}, 1$, and correspond to the
 maximum group velocities of bound-states of 4, 3, 2 and 1
 magnons.} \label{fig:S3logD1}
\end{figure}

The modulus $|z|$ can also be extracted from the $z$ component of the
magnetization, since $M^z=-\sqrt{1-|z|^2}$. Fig.~\ref{fig:S3logD1}
indeed shows an exponential decay which is compatible with the result
of the perturbative expansion.  But importantly, the quantum chain
shows a similar behavior. Although the latter displays additional
oscillations~\cite{bulchandani_subdiffusive_2019}, the tail of the $z$-magnetization profile exhibits some
exponential-like decay in $r/t$ (black and blue lines in
Fig.~\ref{fig:S3logD1} correspond to different times and are almost on
top of each other). This decay is however cut off af $r/t\simeq 1$, because of the maximal velocity in the chain. 
A qualitatively similar behavior is also observed in the XXZ chain for $\Delta>1$ (data not shown).
This exponential decay is however cut for $r/t>1$, which corresponds
to the largest velocity in this lattice model: the maximum group
velocity of an ``up'' spin in a background of down spins is indeed
equal to unity. Similarly, we observe some ``bumps'' in the tail of
the magnetization profile at $r/t\simeq\frac{1}{4}, \frac{1}{3},
\frac{1}{2}$, and these correspond to the maximal group velocities of
bound-states of 4, 3 and 2 magnons propagating in the vacuum ({\it
  i.e.}, down spins) for $\Delta=1$~\cite{denardis_pc_2017}.  Such bumps have
also been observed in the Von Neumann entropy
profiles~\cite{misguich_dynamics_2017}. 
Of course these are lattice
effects and have no analog in the continuum.

\section{Easy axis regime $\Delta>1$}
\label{sec:easy_axis}

In the early study of the XXZ domain wall problem by Gobert {\it et
  al.}~\cite{gobert_real-time_2005}, the current $I(t)\equiv I(r=0,t)$ associated to
the transfer of the $z$-component of the magnetization
(from the left half to the right half of the chain)   
was observed to
vanish at long time for $\Delta>1$. This absence of spin transport is consistent with
exact Bethe ansatz results for the return probability (or Loschmidt
echo)~\cite{mossel_relaxation_2010}. As for the LL problem, the $z$
magnetization profile was observed to freeze at long times
\cite{mallick_2017,gamayun_domain_2019}, and to be described by a
soliton solution of the easy-axis LL equation.
This was explained analytically by the fact that the associated inverse scattering problem involves a stable static kink
in the easy-axis case~\cite{gamayun_domain_2019}.

In this section we show
that, in the limit $\Delta \to 1^+$ both problems have the {\em same}
soliton-like stationary magnetization profiles, with a spatial width
scaling as $\delta^{-1/2}$ and which thus diverges when $\Delta \to 1^+$.
The oscillations about the stationary profile appear to have a period
proportional to $\delta^{-1}$, which can be understood from a
dimensional analysis of easy-axis LL equation.
Finally, as
discussed in Sec.~\ref{sec:sw}, at larger distances the perturbative
analysis still holds and small amplitude oscillations around the $z$
axis propagate much further than the soliton width, but decay
exponentially in $r/t$.

\subsection{Numerics for LL and XXZ}

\begin{figure}[h]\center
\includegraphics[width=0.6\linewidth]{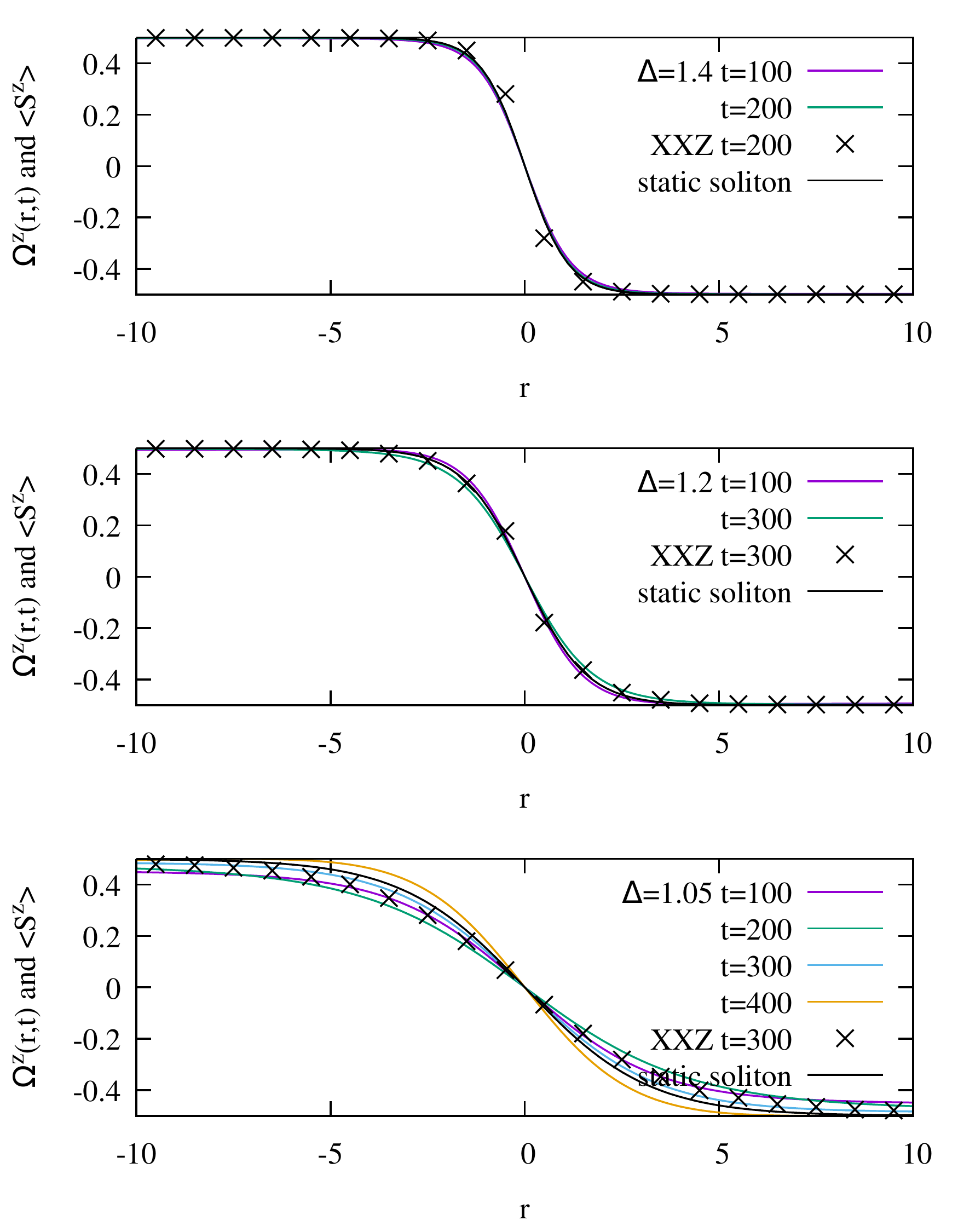}
\caption{ $\Omega^z(r,t)$ and $\langle S^z(r,t)\rangle$ as a function
  of the position $r$ for $\Delta=1.4$ (top), $\Delta=1.2$ (middle)
  and $\Delta=1.05$ (bottom) and for different times from $t=100$ to
  $t=400$. The LL solution shows some long-lived oscillations around
  some mean profile. These are visible (larger amplitude) for
  $\Delta=1.05$.  The full black line represents the static soliton of
  Eqs.~\eqref{eq:static_soliton}.  Black crosses: expectation value
  $\langle \hat S^z \rangle$ in the quantum XXZ chain (tDMRG
  simulations).  To visualize the amplitude and the period of these
  oscillations, see Fig.~\ref{fig:current_Dgt1}.
}\label{fig:SzDelta_gt1}
\end{figure}

The $z$ component of the magnetization profiles is plotted in
Fig.~\ref{fig:SzDelta_gt1} for a few values of $\Delta>1$. The width
of the stationary (or mean) profile grows when approaching the
isotropic point. It nevertheless remains relatively limited even at
$\Delta=1.05$, where it does not exceed 20.  In addition to the
absence of propagation at long times, we observe oscillations with an
amplitude which grows when $\Delta$ approaches 1.  In the panel of
Fig.~\ref{fig:SzDelta_gt1} corresponding to $\Delta=1.05$ we indeed
see that the profile keeps changing up to $t=400$. Remarkably we have
a good quantitative agreement between the quantum chain and LL,
without any adjustable parameter. Again, this should be viewed as a
consequence of the energy conservation argument exposed in
Sec.~\ref{sec:energy}.
The oscillations are easier to see in the current (defined in
Eq.~\eqref{eq:Ib}), plotted in Fig.~\ref{fig:current_Dgt1} for the LL
and XXZ models. In this figure, the left panel shows some XXZ data for
several values of $\Delta$ in the easy-axis regime. The current has
some transient regime, after which it oscillates around zero. When
multiplied by $\sqrt{t}$ (right panel of Fig.~\ref{fig:current_Dgt1})
the amplitude of the oscillations appear to be approximately
constant. The data are thus compatible with a slow $t^{-1/2}$ decay of
the oscillations, at least in the LL case.  The situation for XXZ is
slightly less clear and longer runs would be needed to establish
firmly the form of the decay.

Concerning the temporal period of the oscillations, they appear to be
proportional to $\delta^{-1}$. This can be seen in the middle panel of
Fig.~\ref{fig:current_Dgt1}, where an approximate collapse of the
current oscillations is achieved using a rescaled time $\tau=\delta
t$.  For the XXZ data at different values of $\Delta$, as well as for
an easy-axis LL solution. Here again, we get a reasonable agreement
between the classical and quantum models, without any adjustment.

\subsection{Easy-axis LL soliton} 

The observation made in the previous paragraph can be explained from
the LL point of view, using the rescaled equation
Eq.~\eqref{eq:rescaledLL} and the rescaled time $2\tau = |\delta| t$.
   
One can easily check (see, e.g., Ref. \cite{kosevich_magnetic_1990})
that the following profile is an exact static solution when $\Delta>1$
($\delta>0$):
\begin{subequations}\begin{eqnarray}
2\Omega^x(r)&=& 1  / \cosh \left[  r \sqrt{\delta}  \right] \\
\Omega^y(r)&=&0 \\
2\Omega^z(r)&=&- \tanh \left[ r  \sqrt{\delta}\right].
 \end{eqnarray}\label{eq:static_soliton}\end{subequations}
It is a topological soliton, which lies in the $x$-$z$ plane.  It
corresponds to Eqs.~\eqref{eq:init_cond} with $a=\sqrt{\delta}$, but
this is not the value $a(\Delta)$ which ensures that the energy is
that of the domain wall of the XXZ chain (see
Sec.~\ref{ssec:a_Delta}).  The energy density of this solution is
$\epsilon(r)=\frac{1}{4}\delta\left[\cosh(\sqrt{\delta}x)\right]^{-2}$
and its total energy is $E_S=\frac{1}{2}\sqrt{\delta}$.  This energy is
always lower than the XXZ energy $E_{\rm XXZ}=\frac{1}{2}\Delta$.

\begin{figure}[h]\center
\includegraphics[height=0.35\linewidth]{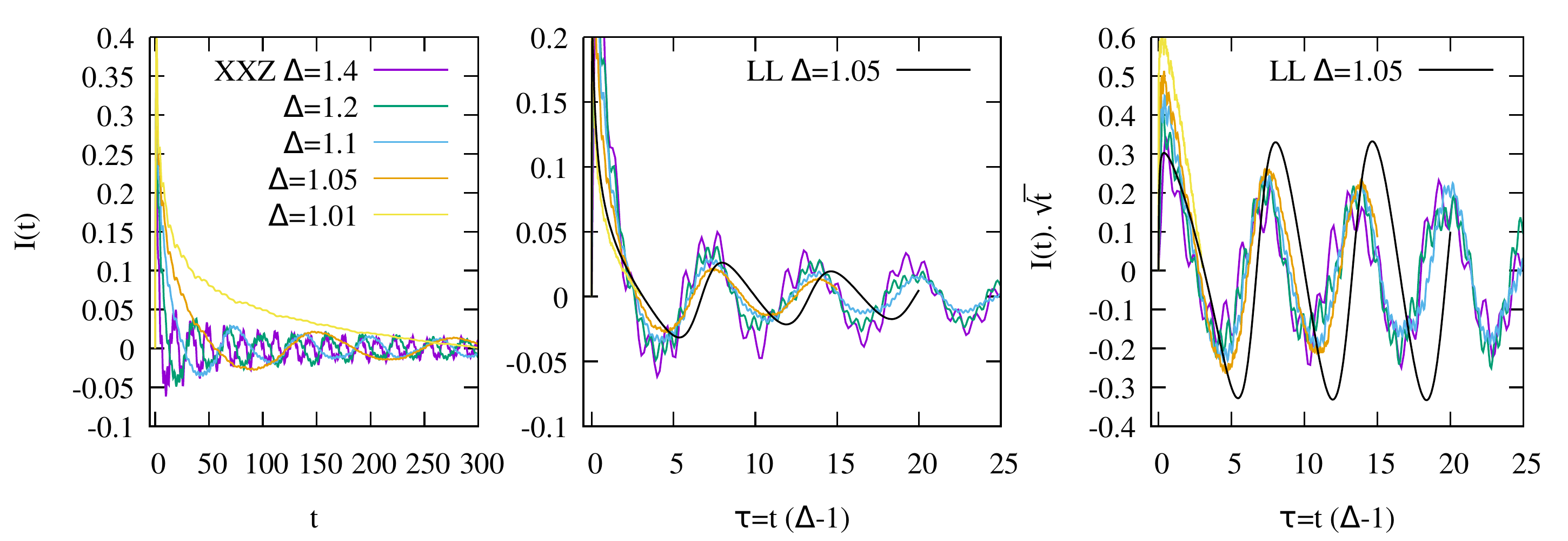}
\caption{Left panel: Current $I(t)$ for different values of $\Delta$,
  from 1.01 to 1.4, in the XXZ chain.  Data obtained from tDMRG
  simulations on a chain with $L=600$ or 800 sites and maximum bond
  dimension between 1000 and 2000.  Middle panel: same data plotted as
  a function of $\tau=(\Delta-1)t$. The relatively good collapse of
  the curves associated to different values of the anisotropy can be
  understood in the framework of the easy-axis LL model, thanks to the
  rescaling defined in Eqs.~\eqref{eq:rescaledLL} which absorbs the
  anisotropy $\delta=2(\Delta-1)$ into a definition of a new time
  variable $\tau=\frac{1}{2}|\delta|t$.  For comparison, the current obtained
  from the LL equation at $\Delta=1.05$ is also shown (black line).
  Right panel: same data multiplied by $\sqrt{t}$, which gives
  oscillations with an approximately constant amplitude.}
 \label{fig:current_Dgt1}
\end{figure}

Nevertheless, the numerical results of
Figs.~\ref{fig:SzDelta_gt1}-\ref{fig:current_Dgt1} show that the LL
and XXZ profiles oscillate around, and then converge to, the soliton
solution above. This convergence becomes very slow as $\Delta$
approaches $1$.  So, as argued in Ref. \cite{gamayun_domain_2019}, we find
that the excess energy is radiated away and that the central region of
the system approaches the static soliton configuration.  The scaling
in Eq.~\eqref{eq:static_soliton} then naturally explains the observed
width of the steady profile, proportional to $\delta^{-1/2}$. The
rescaled time in Eq.~\eqref{eq:rescaledLL} shows that the natural time
scale is $\sim\delta^{-1}$, as the observed period of the oscillations
in $I(t)$, both for LL and XXZ.

\section{Isotropic model $\Delta=1$}\label{sec:Delta=1}

For the quantum chain, the domain wall problem in the isotropic case
has already attracted a lot of attention. Although some
studies~\cite{gobert_real-time_2005,ljubotina_spin_2017} concluded
that the expansion of $\langle S^z\rangle$ could be
superdiffusive, a diffusive behavior with a logarithmic correction now
seems more likely~\cite{stephan_return_2017,misguich_dynamics_2017} (still,
some anomalous diffusion is possible when starting from a different
initial condition that is close to an equilibrium
state~\cite{ljubotina_spin_2017,ilievski_superdiffusion_2018,gopalakrishnan_kinetic_2019,de_nardis_anomalous_2019}).
The same behavior was recently found of the isotropic LL model
\cite{gamayun_domain_2019}, and we provide below a detailed
quantitative comparison of quantum spin chain with the LL model. This
section being relatively long, we begin by a summary of the results.

First, we confirm numerically the findings of
Ref. \cite{gamayun_domain_2019} concerning log-corrected diffusion of
the $z$ component of the magnetization in LL.  But we go further and
provide a detailed analysis of the $x$ and $y$ components, the energy
density, and the energy current in particular.  We find that, both for
the quantum spin chain and the LL problem, two regimes should be
distinguished in the long time limit. First, a region where
$r/\sqrt{t}$ (up to $\ln{t}$ factors) is finite, where the $z$
component is different from $\pm \frac{1}{2}$, and where the energy
density is independent of $r$ and proportional to $\ln(t)/t$. In this
region the magnetization can be described by a self-similar solution
with an effective time-dependent energy density. We also characterize
the regime of $r/t$ finite, which may be called the ballistic
regime. There, $\langle S^z\rangle$ as well as $\Omega^z$ approach
$\pm \frac{1}{2}$ with corrections which scale as $\sim \exp(-cr/t)$.
$r/t$ appears to be the appropriate scaling variable for these
quantities, as well as for the energy or the energy current, contrary
to what the behavior of the $z$ component could naively suggest.  We
also discuss an important difference between the quantum and the LL
problems: the quantum state is invariant under rotation about the $z$
axis (hence $\left<S^x_r\right>=\left<S^y_r\right>=0$), while the
classical state is not. This symmetry difference enables us to
conjecture some logarithmic growth of the entanglement entropy in the
quantum chain (Sec.~\ref{ssec:entanglement}).  We could also find a
counterpart of the $x$ and $y$ components of the LL magnetization in
the quantum side by using two-point correlations
(Sec.~\ref{ssec:correl}).  We also discuss (Sec.~\ref{ssec:filament})
the point of view where the LL magnetization $\vec M$ is the tangent
vector of a curve in the three dimensional space. As an interesting
finding, we observe that the torsion of this curve is remarkably simple
and equal to $ r/t$ at long times, in the diffusive as well as in the
ballistic regime.  We finally explain some of these findings by
exploiting the mapping from the isotropic LL equation to the NLS
equation, and constructing a variational solution to the latter.

\subsection{$z$ component: current and diffusion with logarithmic correction}

Solving numerically the isotropic ($\Delta=1$) Landau-Lifshitz equation
\begin{equation}
 \vec \Omega_t = \vec \Omega \wedge \vec \Omega_{rr}
 \label{eq:LL_Delta=1}
\end{equation}
with the initial condition \eqref{eq:init_cond} reveals some diffusive
behavior for the expansion of the $\Omega^z$ profile, with some
(multiplicative) logarithmic correction.  This confirms the finding of
Ref. \cite{gamayun_domain_2019}.

\begin{figure}[h]\center
\includegraphics[height=0.35\linewidth]{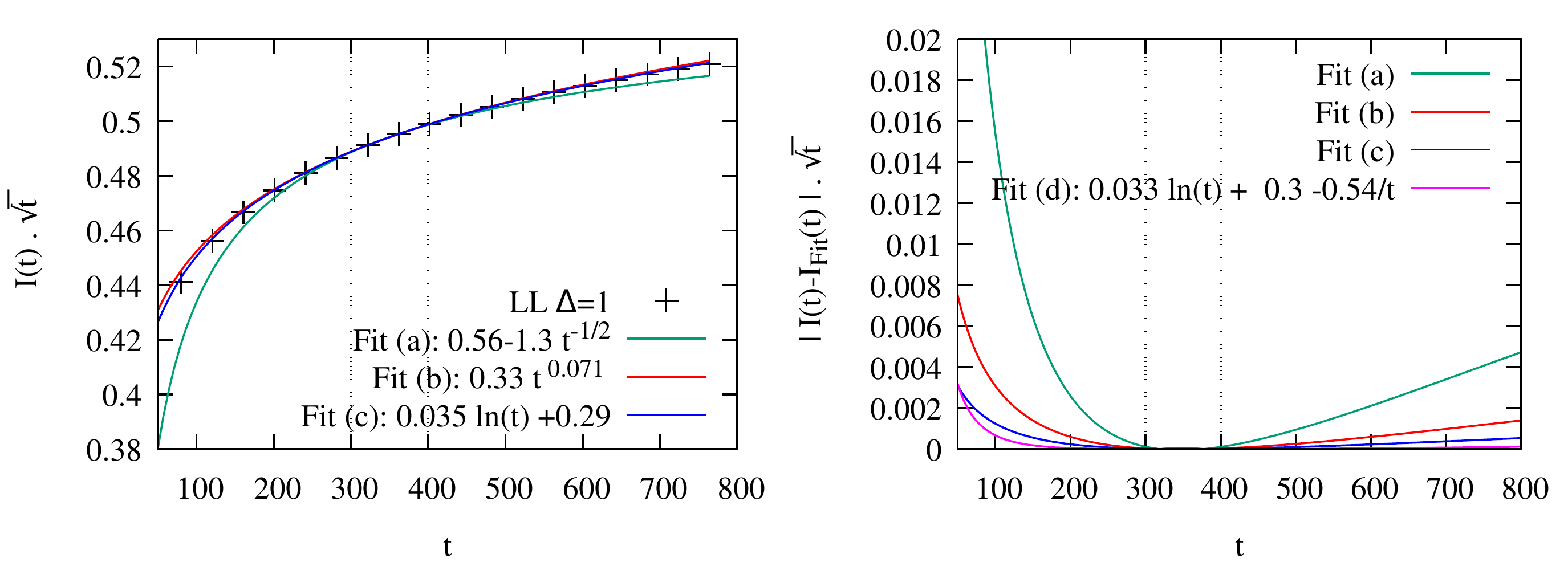}
\caption{Left: LL current $I(t)$ [Eq.~\eqref{eq:Ib}] multiplied by
  $\sqrt{t}$ as a function of time, for $\Delta=1$.  The numerical
  solution is compared with three fitting functions. These fits are
  done using only the interval $[300,400]$ (dotted vertical lines),
  and the quality of these fits can be judged by the discrepancy
  between the data and the fitting functions outside this interval.
  Fit (a) corresponds to $I(t)\simeq \sqrt{t}+{cst}/t$. Fit (b)
  represents a superdiffusive behavior $I(t)\simeq t^\alpha$ with
  $\alpha>0.5$. The third one, (c), corresponding to $I(t)\simeq
  \ln(t)/\sqrt{t}$, appears to provide the best fit, and is in
  agreement with the results of Ref. \cite{gamayun_domain_2019}.  The
  right panel shows the absolute value of the difference between the
  LL solution and the three fitting functions. Among the three fits
  (a),(b) and (c), the fit (c), which corresponds to a diffusive
  behavior with a multiplicative logarithm shows the best agreement
  with the numerical solution of the LL equation.  Fit (d): including
  an additional $1/t$ correction to the function of fit (c) makes the
  agreement almost perfect at the scale of this plot.
} \label{fig:FigCurrentSQRT_D=1}
\end{figure}

\begin{figure}[h]\center\includegraphics[width=0.65\linewidth]{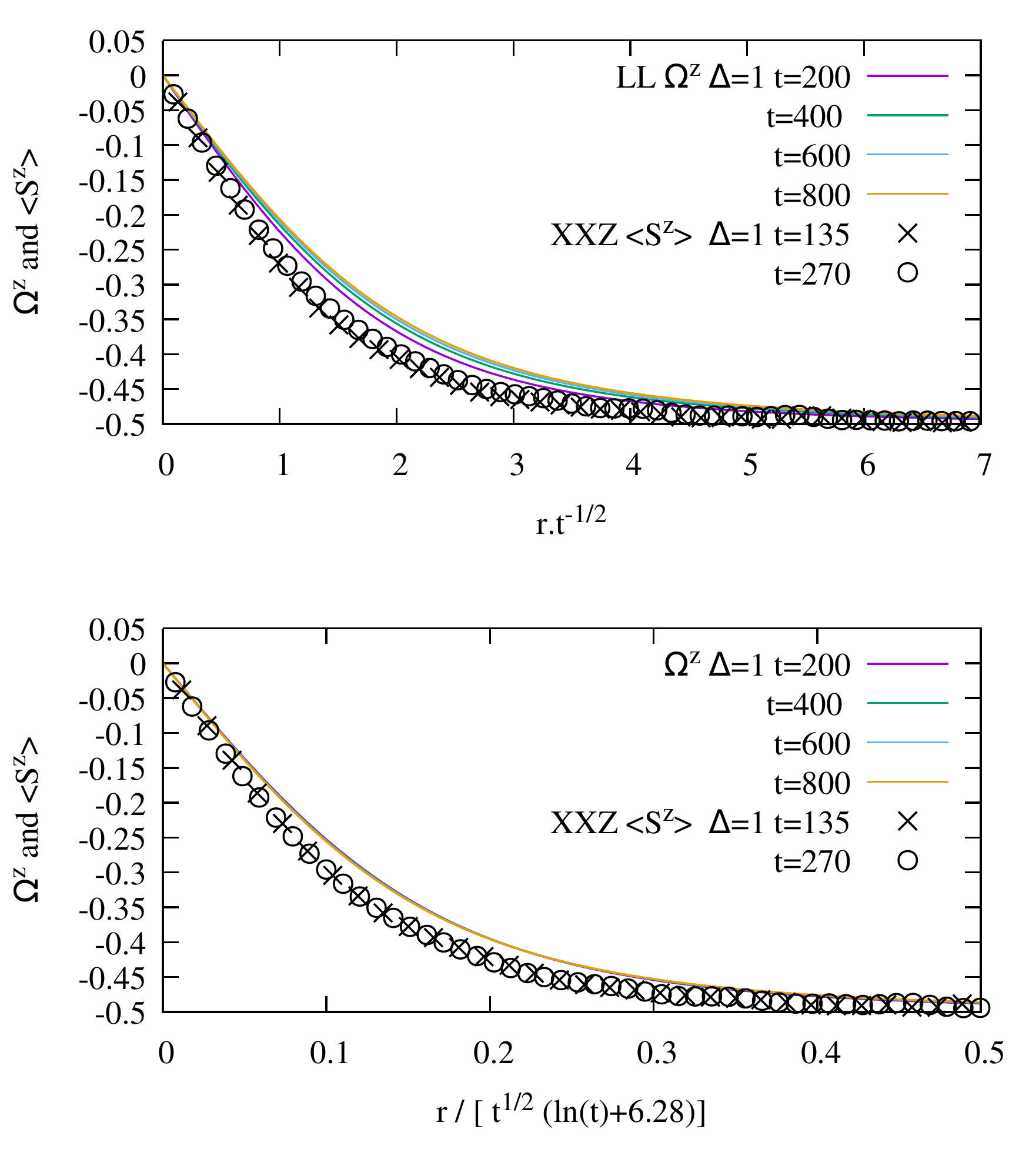}
  \caption{$\Omega^z$ magnetization profile for LL at $\Delta=1$, and
    comparison with $\langle S^z\rangle$ in the XXZ chain (tDMRG
    simulations $L=800$ sites, maximal bond dimension $\chi=2000$ and
    time $t=270$). The horizontal axis is $r/\sqrt{t}$ in the top
    panel, and $r/g(t)$ with $g(t)=\sqrt{t}\left[\ln(t)+6.28\right]$
    in the lower panel. The latter, gives a good collapse of the
    curves obtained at different time.  The same multiplicative
    logarithmic correction to a diffusive behavior was also observed
    in Ref.~\cite{misguich_dynamics_2017} for the XXZ chain and in
    Ref. \cite{gamayun_domain_2019} for LL.  Remark: the constant 6.28 in
    $g(t)$ is equivalent to the function (c) in
    Fig.~\ref{fig:FigCurrentSQRT_D=1}, since
    $0.07\frac{d}{dt}\left[(\ln(t)+6.28)\sqrt{t}\right]=\left[0.29+0.035\ln(t)\right]/\sqrt{t}$.
  }
 \label{fig:Profile_D=1}
\end{figure}

This can be seen by analyzing the current $I(t)$, as summarized in
Figs.~\ref{fig:FigCurrentSQRT_D=1} and \ref{fig:Profile_D=1}.  For a
conventional diffusive behavior, $I(t)$ would decrease as
$t^{-\frac{1}{2}}$, and $I(t) \sqrt{t}$ would converge to a constant
at long times.  We instead see (left panel of
Fig.~\ref{fig:FigCurrentSQRT_D=1}) that it does not saturate, at least
up to $t=800$.  The data are then fitted to (a) a constant minus a
$1/\sqrt{t}$ correction, (b) a power low ({\it i.e.} superdiffusion)
and (c) to a logarithm.  The latter offers the best fit, and agrees
with the LL solution far beyond the time interval $[300,400]$ where
the fit was performed (interval marked by the vertical dotted
lines).\footnote{Note that (c) is equivalent to Eq. (7) in
  Ref.~\cite{misguich_dynamics_2017}, and was advocated to describe
  the behavior of the domain wall problem in the quantum Heisenberg
  chain.}  As can be seen in the right panel of
Fig.~\ref{fig:FigCurrentSQRT_D=1}, the agreement can be further
improved by adding an additional $1/t$ correction [curve (d)].

Since the form $I(t)\sim \ln(t)/\sqrt{t}$ offers the best match with
the data, this Ansatz for the current can be integrated over time to
give the transfered magnetization from the left to the right since the
initial time, and the result is $g(t)\simeq\sqrt{t}(\ln(t)+a)$. As
shown in the bottom panel of Fig.~\ref{fig:Profile_D=1}, once the
magnetic profiles $\Omega^z(r,t)$ and $\left<S^z_r(t)\right>$ are
plotted as a function of $r/g(t)$, the data obtained at different
times collapse onto the same curve. The collapse is in particular much
better than if one simply plots $\left<S^z_r(t)\right>$ (or $\Omega^z(r,t)$) as a function of $r/\sqrt{t}$,
as done in the upper panel of Fig.~\ref{fig:Profile_D=1}.

The stationary profiles obtained in the bottom of
Fig.~\ref{fig:Profile_D=1} appear to be slightly different for the LL
problem and the quantum spin chain. From the conjectured asymptotic
equivalence of the two problems one could have expected to see the
very same profile in the two cases. But instead we may attribute the
discrepancy to some memory of the initial state and of the short time
evolution, where the energy density is finite and the two problems
still inequivalent.

\subsection{$U(1)$ symmetry and entanglement entropy in the XXZ chain}
\label{ssec:entanglement}

The states described by the LL equation break the rotational
invariance about the $z$ axis. This is particularly obvious for the
initial condition \eqref{eq:init_cond} where
$\Omega^x(r=0,t=0)=\frac{1}{2}$ and $\Omega^y(r,t=0)=0$. On the other
hand, the domain wall state used as an initial condition for the
quantum chain is invariant under rotations about the $z$ axis, and the
state $|\psi(t)\rangle$ therefore remains an eigenstate of $\hat
S^z_{\rm tot}=\sum_r \hat S^z_r$ at any time.

We may thus take the classical LL state, and the associated
magnetization vectors for all the integer values of the position $r$,
and promote it to a (quantum) product state $|\psi(t)_{\rm LL}\rangle$
for the spin chain. In order to restore the $U(1)$ symmetry, we can
rotate it (globally) about the $z$ axis, and perform a linear
combination of different rotated copies of this state.  The resulting
state could then be written a $|\psi(t)_{\rm LL,
  sym}\rangle=\int_0^{2\pi}d\theta \exp\left(i\theta\hat S^z_{\rm
    tot}\right) |\psi(t)_{\rm LL}\rangle$
\cite{castro-alvaredo_entanglement_2012,misguich_finite-size_2017}.
It is then clear that this $U(1)$-symmetric state is no longer a
product state. The symmetrization thus introduced some quantum
entanglement in the system.

How can we estimate the Von Neumann entropy associated to a left-right
partition on the chain ?  We should first estimate the number of
linearly independent states which are generated by the rotations
above. For a macroscopic quantum spin of large size $S$, the rotations
would span a space of dimension $\mathcal{O}(S)$
\cite{castro-alvaredo_entanglement_2012,misguich_finite-size_2017}. We
may use a similar argument for $|\psi(t)_{\rm LL}\rangle$, which we assume here to
be nontrivial only over a spatial region $[-R(t),R(t)]$, and which is
therefore a state involving different spin quantum numbers up to
$S_{\rm max}\sim\mathcal{O}(R(t))$. We may thus consider that the
symmetrized state $|\psi(t)_{\rm LL, sym}\rangle$ is actually a linear
combination of $\mathcal{O}(R(t))$ states which are orthogonal to each
other. In turn, we may take the log of this dimension to estimate the
von Neumann entropy. For large $R(t)$ we would thus have $S_{\rm
  vN}\sim \ln(R(t))$.

\begin{figure}[H]\center\includegraphics[width=0.65\linewidth]{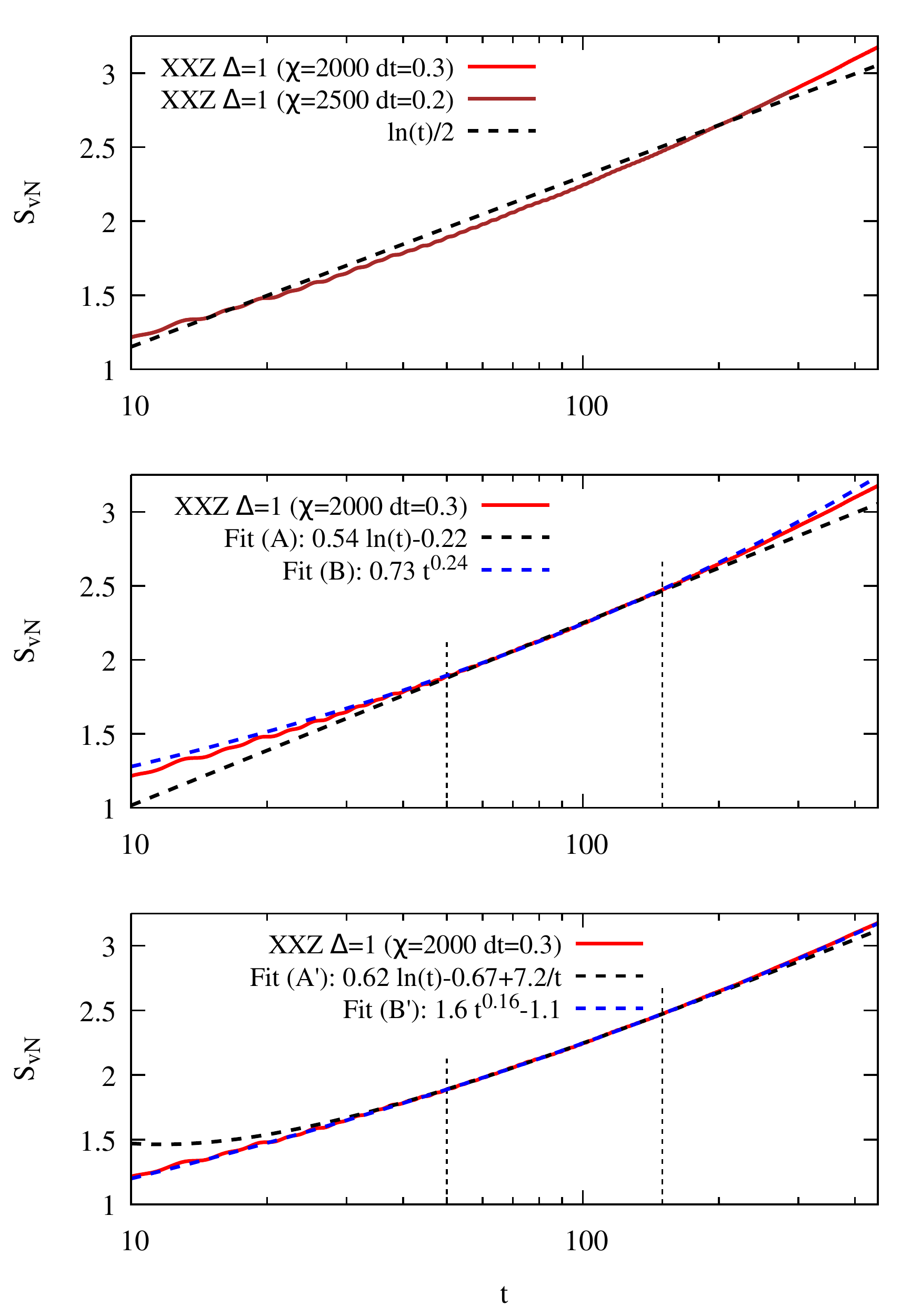}
 \caption{Von Neumann entanglement entropy for the quantum model at $\Delta=1$, for a partition in the center of the chain.
 Top: 
 tDMRG calculations with $L=800$, time step $dt=0.2$, maximum bond dimension $\chi=2500$, MPS truncation parameter $10^{-11}$, with time up $t=286$ (brown line).
  At the scale of the plot the results are unchanged when taking $dt=0.3$, $\chi=2000$ and the truncation parameter to $10^{-10}$  (data shown up to $t=400$, red line), which indicates a good convergence. We also show $\frac{1}{2}\ln(t)$ (see Sec.~\ref{ssec:entanglement}), which however underestimates the entropy growth at large times.
 Middle: two-parameter fits of the entanglement entropy, with (A)  $a\ln(t)+b$ or (B) a power law $c\,t^d$.
 Bottom: three-parameter fits of the entanglement entropy, with (A')  $a\ln(t)+b+c/t$ or (B') $c\,t^d+e$.
 The fit (B'), with an exponent close to $1/6$, seems to offer the best description of the data, but a logarithmic growth cannot be excluded.
 All the fits are performed in the interval $[50,150]$ (dashed vertical lines).
  }\label{fig:SvN}
\end{figure}

We plot in Fig.~\ref{fig:SvN} the entanglement entropy up to $t=400$
at $\Delta=1$, compared with $\frac{1}{2}\ln(t)$. The latter would be
the leading term in the entropy generated by symmetrization for a
region of length $\sim \sqrt{t}$, corresponding to diffusion.\footnote{
The present argument does not apply away from $\Delta=1$. In particular for $\Delta<1$ the
nearest-neighbor correlations deviate from
those of a perfectly ferromagnetic state (by an amount which is $\mathcal{O}(|\delta|)$, see Sec.~\ref{sec:energy}), even at long times.
 Although these  small deviations may only
weakly affect  the local magnetization  or the energy density,  they
change completely the entanglement  entropy. 
 This can be illustrated,
for   instance,  by   considering  a   much simpler  case:  the
homogeneous ground-state of the xx chain,  with a magnetization per site $m$ close
(but not  equal) to  $-1/2$.  The spin-spin  correlations can  be made
arbitrarily close  to that of a  ferromagnet by taking  $m$ to $-1/2$.
Nevertheless the  quantum state  is a (low  but finite  density) Fermi
sea.  In such a case the entropy of a long segment of length $l$ scales
as $\sim \log(l)/3$.  This is clearly very different from the situation with $m=-1/2$
exactly, which is a product state.
Going back to the domain wall problem, the free fermion case $\Delta=0$ indeed gives $S_{\rm vN}\sim\frac{1}{6}\ln(t)$~\cite{eisler_surface_2014,dubail_conformal_2017}.
As for tDMRG results (data not shown) at $\Delta=0.5$ and $0.7$ (up to $t\sim 200$)
suggest that this scaling may hold for all $\Delta<1$.}
As already mentioned in Ref.~\cite{misguich_dynamics_2017}, the
  entanglement entropy data appear to be compatible with $\sim\ln(t)$,
but, with the times accessible to our simulations, the coefficient of
the logarithm seems slightly larger than 0.5 (see fits (A) and (A') in Fig.~\ref{fig:SvN}). But we should stress that a power-law behavior
can also not be excluded~\cite{ljubotina_spin_2017,misguich_dynamics_2017}.
In particular, the present data appear to be well described by 
$S_{vN}\sim t^{0.16}$, as shown by the fit (B') in the bottom panel
of Fig.~\ref{fig:SvN}.
In any case, $\frac{1}{2}\ln(t)$ is only the semiclassical contribution to the entanglement, and
it appears that some other contribution(s) to the entanglement
entropy need to be included, like for instance that carried by the ballistic magnons (and bound states
of magnons) in the region where $r/t$ is finite.  We leave for future
studies the task of determining the precise law of entanglement growth
in this problem at $\Delta=1$, but the $U(1)$ symmetry argument above
should provide a starting point.

\subsection{Correlations in the quantum chain and $x$ and $y$ components of the LL magnetization}
\label{ssec:correl}
As explained above, there is a $U(1)$ symmetry in the
spin-$\frac{1}{2}$ chain, and this implies that $\langle \hat
S^x_r\rangle=\langle \hat S^y_r \rangle=0$ at any time. On the other
hand, the states described by the LL equation break the rotational
invariance about the $z$ axis.

It is nevertheless possible to compare $\Omega^x$ or $\Omega^y$
computed in the LL framework with data from the quantum chain. To do
so, one has to consider the two-point spin-spin correlations in the
quantum chain, like $\langle \hat S^x_0 \hat S^x_r\rangle$.  One can
obviously write $\hat S^x_0=\hat P_x-\frac{1}{2}$, where $\hat
P_x=\hat S^x_0+\frac{1}{2}$ projects onto states where
$S^x_0=\frac{1}{2}$.  Thanks to the rotational invariance mentioned
above, we have $\langle \hat S^x_0 \hat S^x_r\rangle=\langle \hat
P^x_0 \hat S^x_r\rangle$.  Since $P^x_0=1$ with probability
$\frac{1}{2}$ (again due to rotational invariance), and zero
otherwise, it is natural to compare $\langle \hat S^x_0 \hat
S^x_r\rangle$ with $\Omega^x$ in the LL problem, where, by
construction, $\Omega^x(r=0,t)=1/2$.  Concerning the normalization,
consider the ferromagnetic state of $N$ spins $1/2$ that has $S^z_{\rm
  tot}=0$ (but total spin $S_{\rm tot}=N/2$). Such a state is the
symmetrized version of the ferromagnetic state pointing in the (say)
$x$ direction.  In such state one can easily compute the spin-spin
correlations: $\langle \hat S^x_0 \hat S^x_r\rangle=\langle \hat S^y_0
\hat S^y_r\rangle=\frac{N}{8(N-1)}$ for $r\ne0$, which approaches
$\frac{1}{8}$ for a large system. We should then compare $4\langle
\hat S^x_0 \hat S^x_r\rangle$ with $\Omega^x(r,t)$, and similarly,
$4\langle \hat S^x_0 \hat S^y_r\rangle$ with $\Omega^y(r,t)$.

\begin{figure}[h]\center
\includegraphics[width=0.6\linewidth]{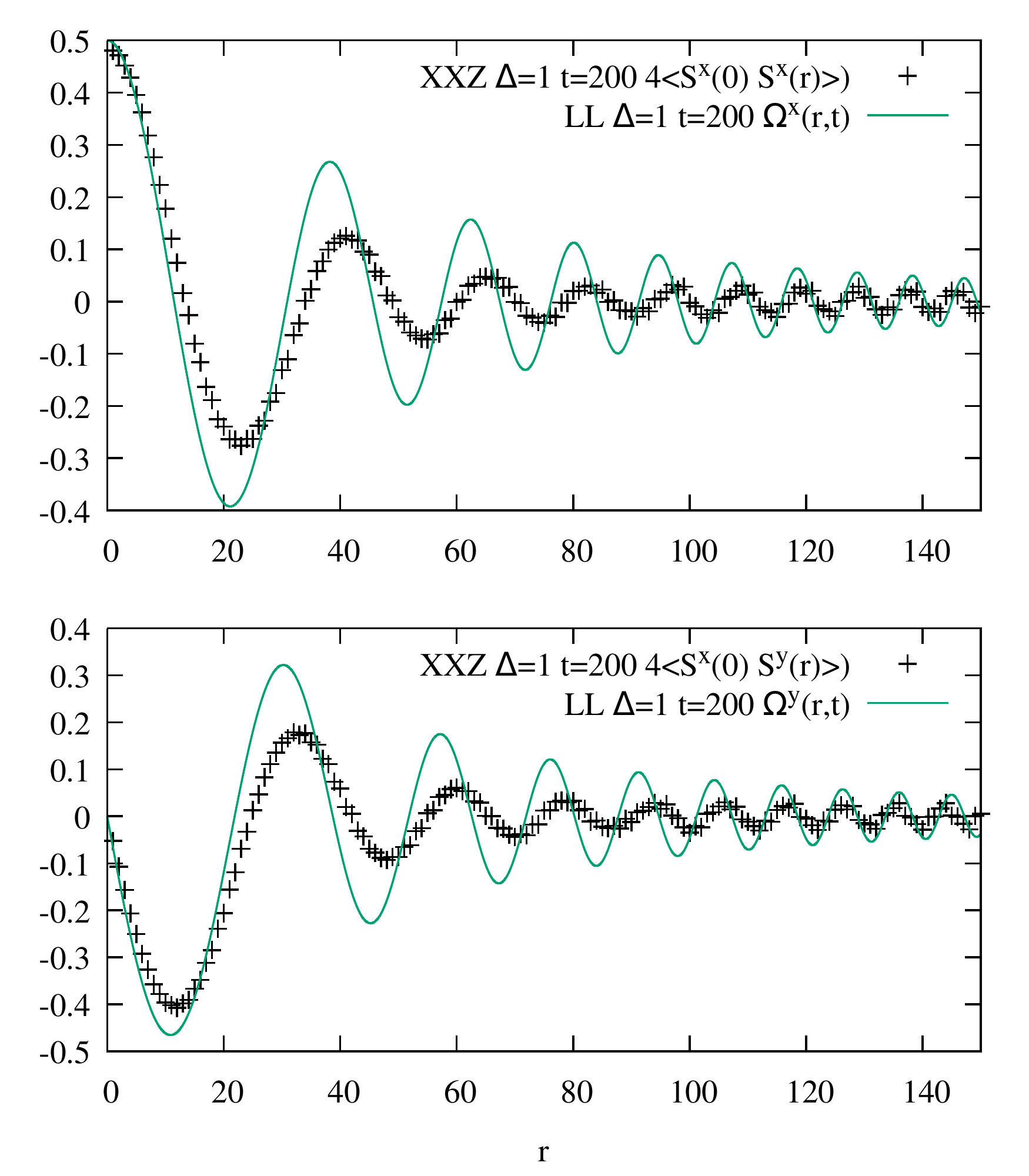}
\caption{Comparison between some two-point correlations in the
  spin-$\frac{1}{2}$ chain with $\Omega^{x,y}(r,t)$ in
  the isotropic LL problem. Upper panel: $\Omega^x$ versus $\langle
  \hat S^x(0)\hat S^x(r\ne0)\rangle$.  Bottom panel: $\Omega^y$ versus
  $\langle \hat S^x(0)\hat S^y(r\ne0)\rangle$.  Both taken at time $t=200$.
  See Sec.~\ref{ssec:correl}.
}\label{fig:XXXcorrel}
\end{figure}

The result is displayed in Fig.~\ref{fig:XXXcorrel}.  First we see
in the upper panel that $4\langle \hat S^x_0 \hat S^x_r\rangle$ approaches $\frac{1}{2}$
when $r\to0$, in agreement with the argument above and the fact that,
in the center of the system and at long times, the quantum chain
resembles locally a symmetrized ferromagnetic state with
$\left<S^z_r\right>=0$.  For larger $r$ we observe some some
semi-quantitative agreement between the quantum and classical models,
without any adjustable parameter.  In particular, the period of the
spatial oscillations in the correlations of the quantum systems is
quite close to the period of the oscillations in the LL magnetization.
An even better agreement between correlations in the
spin-$\frac{1}{2}$ chain and in-plane magnetization of the LL problem
is shown in the bottom panel of Fig.~\ref{fig:SxDelta_lt1} in an
easy-plane case.  The amplitude of the correlations in
Fig.~\ref{fig:XXXcorrel} appears however to decay faster than the LL
magnetization.  We note that this may be due in part to the fact, for
$r>0$, the $\langle \hat S^z\rangle$ profile lies slightly below
$\Omega^z$, as can be seen in Fig.~\ref{fig:Profile_D=1}.  Another
source of discrepancy is the fact that, in the quantum chain, the
center of symmetry is in the middle of a bond, at $r=1/2$, while it is
at $r=0$ in the LL model.  Finally, it is plausible that some quantum
effects remain here for this observable, even at long times.

For $r/t$ finite and $\Delta=1$, the $z$ component of the magnetization tends to $\pm\frac{1}{2}$, both
in the quantum and classical problems. This is a regime where the expansion
of Sec.~\ref{sec:sw} applies and the oscillations of the $x$ and $y$ components of the
magnetization in LL are described by
an azimuthal angle $\varphi (r,t)=-\frac{r^2}{2t}$. This regime is realized in the right part of Fig.~\ref{fig:XXXcorrel}.
On the other and, for $r$ of the order of $\sqrt{t}$ (or even $\sqrt{t}\ln
t$), which corresponds to the first few oscillations in the left of Fig.~\ref{fig:XXXcorrel}, a different behavior of $\varphi$ is observed, as
discussed in the next paragraph.  

\subsection{Energy density}\label{ssec:energy_density}

\begin{figure}[h]\center\includegraphics[height=0.35\linewidth]{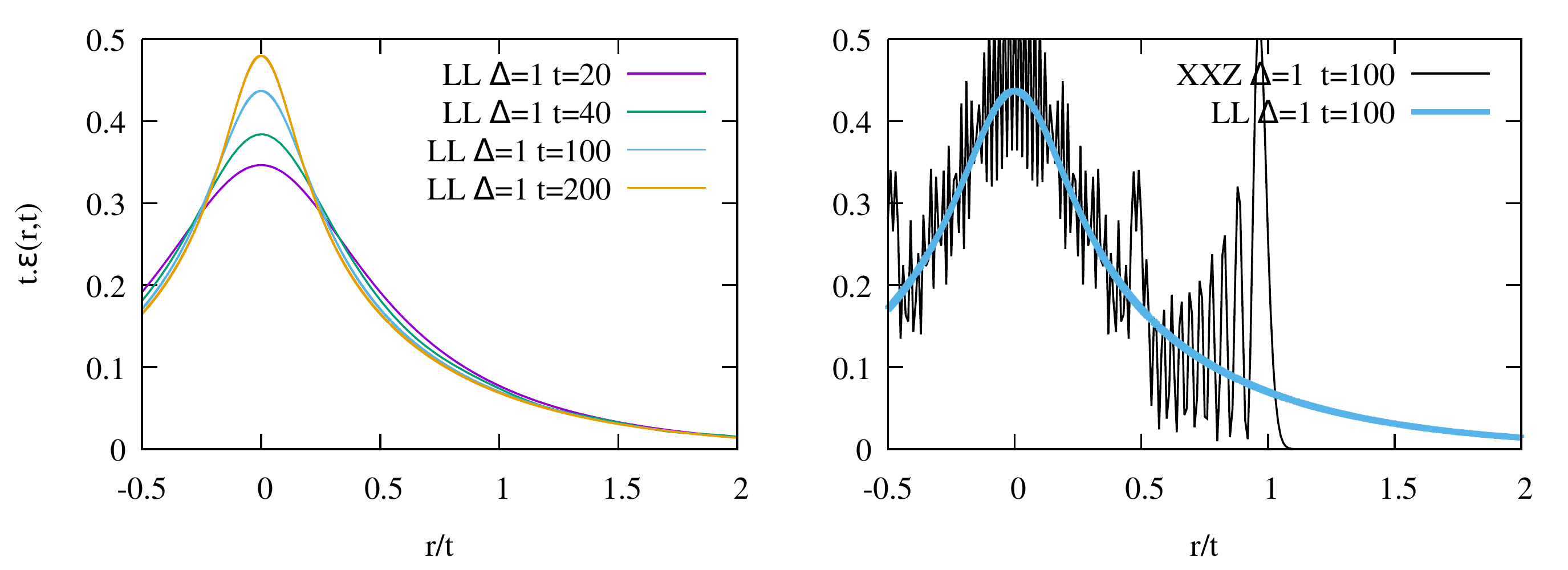}
  \caption{Left panel: LL energy density $\epsilon(r,t)=
    \frac{1}{2}\left(\vec \Omega_r\right)^2$ multiplied by $t$, as a
    function of $r/t$ for $\delta=0$ (isotropic model).  At sufficient
    large $r/t$, the curves associated to different times are
    practically on top of each other. The growth of $\epsilon(r=0,t)$
    with $t$ is compatible with a logarithmic time dependence ($\sim
    \ln t$ behavior, see Fig.~\ref{fig:Energy_density_r=0}).
    Right panel: The LL results are compared with XXZ data (black
    line) at $t=100$ (longer times give a similar curves). For the
    latter there is a maximum group velocity equal to one, which
    explains why the ``signal'' is essentially zero for
    $|r/t|>1$. Thanks to the energy conservation, the area of all
    these curves (LL and XXZ) is equal to $\frac{1}{2}$.
  }\label{fig:FigEN_D=1}
\end{figure}

\begin{figure}[h]\center\includegraphics[height=0.35\linewidth]{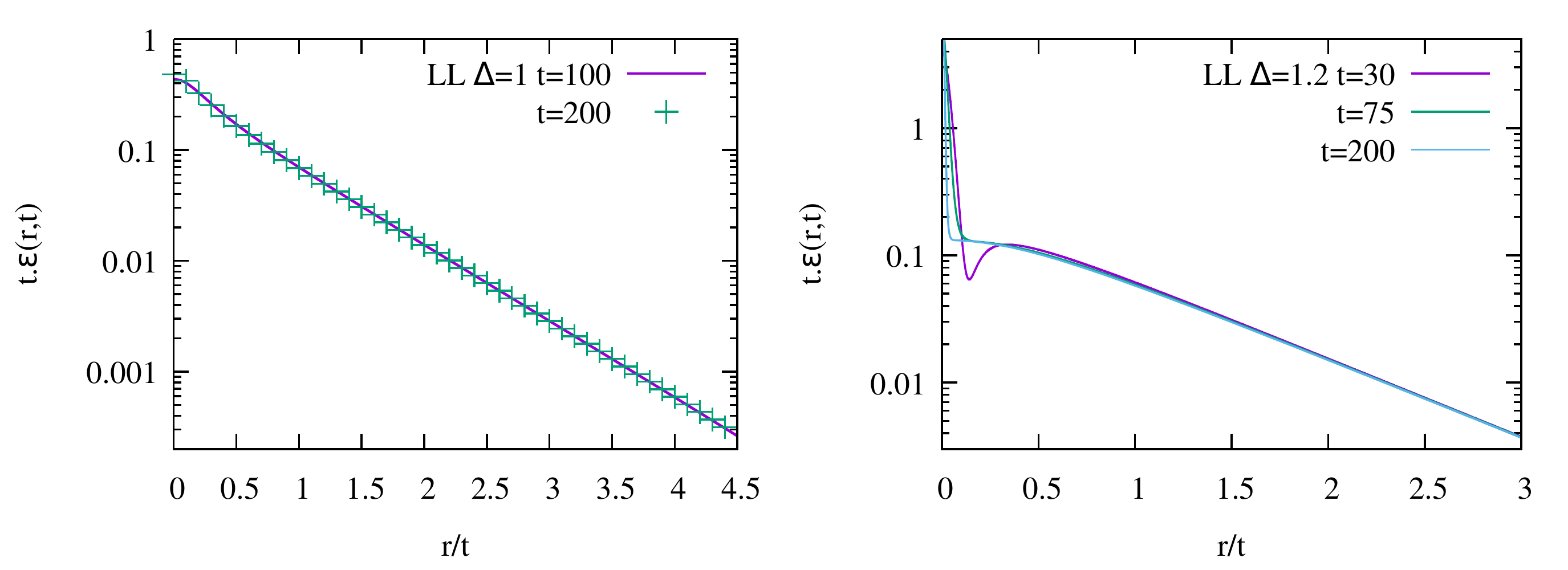}
\caption{Left panel: LL energy density multiplied by $t$ in the isotropic case ($\Delta=1$).
  Away from the center the energy density appears to decay as
  exponentially, $\epsilon(r,t)t \sim \exp(-c r/t)$ with $c\simeq
  1.6$. See Sec.~\ref{sec:var_NLS} for an explanation of this behavior
  (variational approach).  Right panel: LL energy density multiplied
  by $t$ in an easy-axis case ($\Delta=1.2$). At long times we again
  have $\epsilon(r,t)t \sim \exp(-c' r/t)$ for $r/t>0$, but a
  different behavior for $|r|\ll t$ (see Sec.~\ref{sec:easy_axis}).
}\label{fig:FigENlog_D=1}
\end{figure}

\begin{figure}[h]\center
\includegraphics[width=0.6\linewidth]{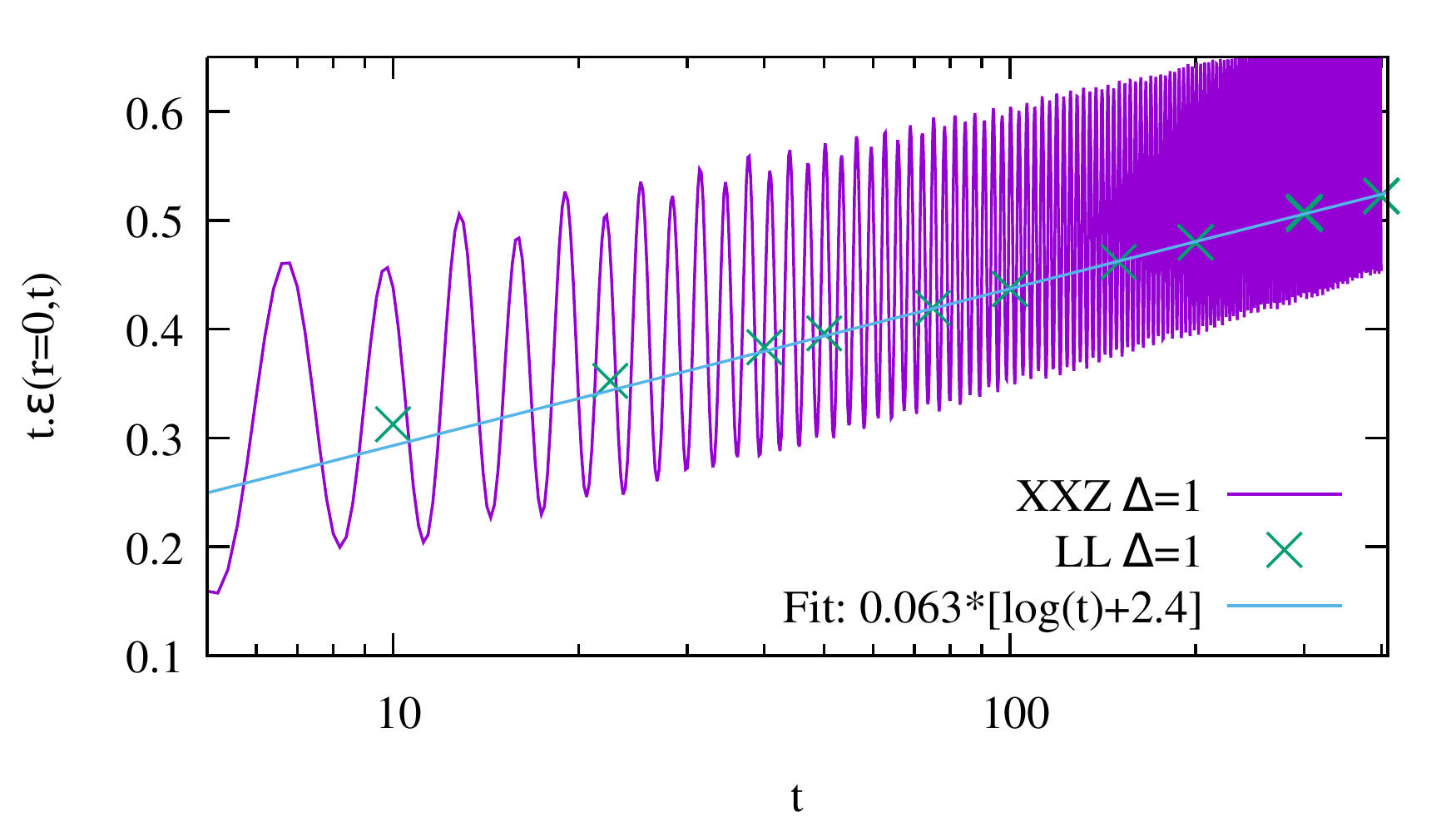}
\caption{Energy density $\epsilon(r=0,t)$ in the center of the
  system and multiplied by $t$, plotted as a function of $t$ (logarithmic scale).
  The green crosses correspond to the LL model and the purple curve corresponds to
  the quantum chain (both at $\Delta=1$). The data are compatible with a logarithmic
  increase with time (see fitting function in the legend, blue line).}\label{fig:Energy_density_r=0}
\end{figure}

For the isotropic LL model the energy density [Eq.~\eqref{eq:LL-H}] is
$\epsilon(r,t)=\frac{1}{2}\left( \vec \Omega_r\right)^2$.  This
quantity is displayed in Figs.~\ref{fig:FigEN_D=1} and
\ref{fig:FigENlog_D=1}. The top panel of Fig.~\ref{fig:FigEN_D=1}
shows that, at least for sufficiently large $r/t$, the energy density
multiplied by time is well described by a function of $r/t$:
$\epsilon(r,t)=\frac{1}{t}K(r/t)$. For small $r/t$, the collapse of
the energy curves associated to different values of $t$ is not good,
and this indicates that $K$ is singular at the origin.  The data
plotted in Fig.~\ref{fig:Energy_density_r=0} indicate a 
logarithmic divergence of $t\cdot\epsilon(r=0,t)$ with time.  From this
we may infer that (in the limit of infinite time) the function $K$
exhibits a logarithmic divergence: $K(u)\sim |\ln u|$ at small
$u=r/t$.

The energy density of the spin-$\frac{1}{2}$ Heisenberg model at $r=0$, measured 
through $\left<\frac{1}{4}-\vec S_0\cdot\vec S_1\right>$ on the central bond of the chain, is also represented in Fig.~\ref{fig:Energy_density_r=0},
as a function of time. It displays oscillations, but otherwise follows relatively well the LL behavior,
namely $\epsilon(r=0)\sim 0.06\left[\ln(t)+2.4\right]/t$ (see the fitting function in Fig.~\ref{fig:Energy_density_r=0}).

In the right panel of Fig.~\ref{fig:FigEN_D=1} the spatial variations of the energy density for
LL is compared to those of the Heisenberg chain. Although the energy
density of the quantum chain shows some important oscillations (their scaling
in the vicinity or $r/t=1$ is discussed in \cite{bulchandani_subdiffusive_2019}), it
follows the LL data relatively well, and it certainly extends beyond
the diffusive scale. It should be noted that there is again no adjustable
parameter here, except for the fact that the width the of LL profile
at $t=0$ was chosen to ensure a total energy equal to that of the
quantum chain (Sec.~\ref{ssec:a_Delta}).

After averaging the short-distance oscillations, this energy density
shows a maximum in $r/t\simeq 1$ and $r/2\simeq 0.5$.  As observed
already in Fig.~\ref{fig:S3logD1}, these two velocities correspond
respectively to the maximum group velocity of a single magnon, and to
the maximum group velocity of a bound state of two-magnons. The
effects of these discrete velocities are specific to the quantum chain
and have no direct analog in the LL problem.

For the isotropic LL model, the energy density can be written using
the polar angles [Eq.~\eqref{polarization}] and their space
derivatives:
\begin{equation}
\epsilon=\frac{1}{2}\left(\vec \Omega_r\right)^2=
\frac{1}{8}\left(\vec M_r\right)^2=\frac{1}{8}\left[(\theta_r)^2
+(\sin\theta \;\varphi_r)^2 \right].\label{eq:epsilon_theta_phi}
\end{equation}
From Fig.~\ref{fig:Profile_D=1} (bottom panel) we know that the
spatial scale for the variation of $\Omega^z$ is $\sim \sqrt{t} \ln
t$.  As a consequence, the gradient $\theta_r$ must scale as $\sim
\left(\sqrt{t} \ln t \right)^{-1}$ in this region.  So, the observed
logarithmic divergence in $t\cdot\epsilon(r=0,t)$ is not associated to
the $(\theta_r)^2$ term, but must come from the second term in
Eq.~\eqref{eq:epsilon_theta_phi}.  We conclude that the gradient of
the azimuthal angle should scale as $|{\varphi_r}_{|r=0}| \sim
\sqrt{\ln{t}/t}$.  Numerically we indeed find that $|\varphi| \simeq
0.22\pi r \sqrt{(\ln{t}+2.5)/t}$ fits the data relatively well for
small $r/\sqrt{t}$, corresponding roughly to one full rotation about
the $z$ axis ($-\pi <\varphi<\pi$).

\begin{figure}[h]\center\includegraphics[height=0.35\linewidth]{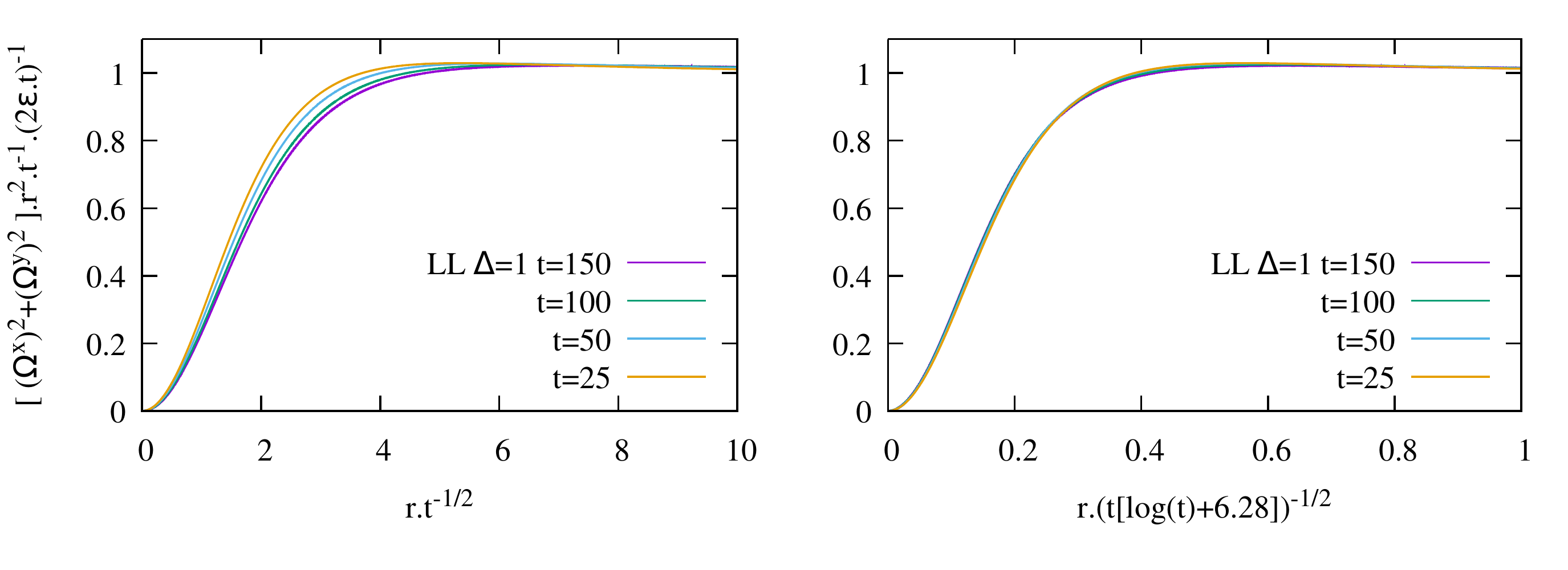}
  \caption{Ratio of the energy density computed in the perturbative
    approximation to the actual energy density.  See text and
    Eq.~\eqref{eq:epsilon_sw} for details. The good curve collapse
    observed in the right panel indicates that this ratio obeys a
    scaling form similar to that of $\Omega^z$ (compare with the
    bottom panel of Fig.~\ref{fig:Profile_D=1}).  }
  \label{fig:R}
\end{figure}

In the limit where $\Omega^z$ is close to $\pm\frac{1}{2}$ the energy
density given in Eq.~\eqref{eq:epsilon_theta_phi} is dominated by the
term proportional to $(\varphi_r)^2$.  From the perturbative expansion of
Sec.~\ref{sec:sw} we have $\varphi=-r^2/(2t)+t\delta/2$ and
$\varphi_r=-r/t$.  Since, by definition,
$\sin(\theta)=2\sqrt{(\Omega^x)^2+(\Omega^y)^2}$, we can write the
energy density in the perturbative approximation as:
\begin{equation}
 \epsilon_{\rm SW}=\frac{1}{2}
\left[(\Omega^x)^2+(\Omega^y)^2\right] \left(r/t\right)^2
 \label{eq:epsilon_sw}
\end{equation}
In other words, the ratio
\begin{equation}
 \epsilon_{\rm SW}/\epsilon=
\frac{\left[(\Omega^x)^2+(\Omega^y)^2\right] r^2}{ 2t^2\epsilon }
 \end{equation}
 should approach $1$ when the perturbative approximation becomes
 asymptotically exact.  This quantity is plotted in
 Fig.~\ref{fig:R}. As expected, the length scale beyond which the
 perturbative calculation becomes accurate is $\sim \sqrt{t\ln t}$, since
 this scale governs how $\Omega^z$ approaches $\pm\frac{1}{2}$.

\subsection{Vortex filament, torsion and self-similar solutions}
\label{ssec:filament}

\begin{figure}[h]\center\includegraphics[height=0.35\linewidth]{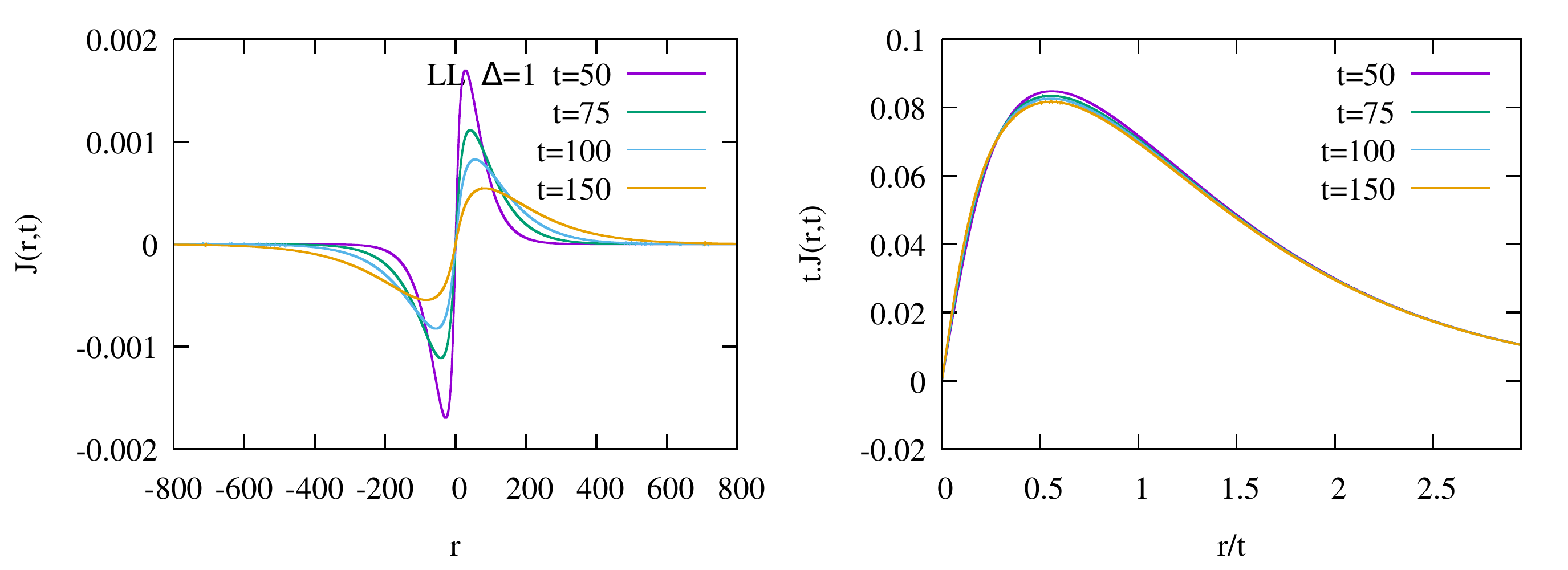}
 \caption{Left panel: LL energy current $J=\vec \Omega \cdot \left(
     \vec \Omega_r \times \vec \Omega_{rr}\right)$ at $\Delta=1$,
   plotted as a function of the position $r$ for different times.
   Right panel: same data, multiplied by $t$ and plotted as a function
   $r/t$. An excellent collapse is observed for $r/t\gtrsim1.5$.  }
 \label{fig:J}
\end{figure}

\begin{figure}[h]\center
\includegraphics[width=0.6\linewidth]{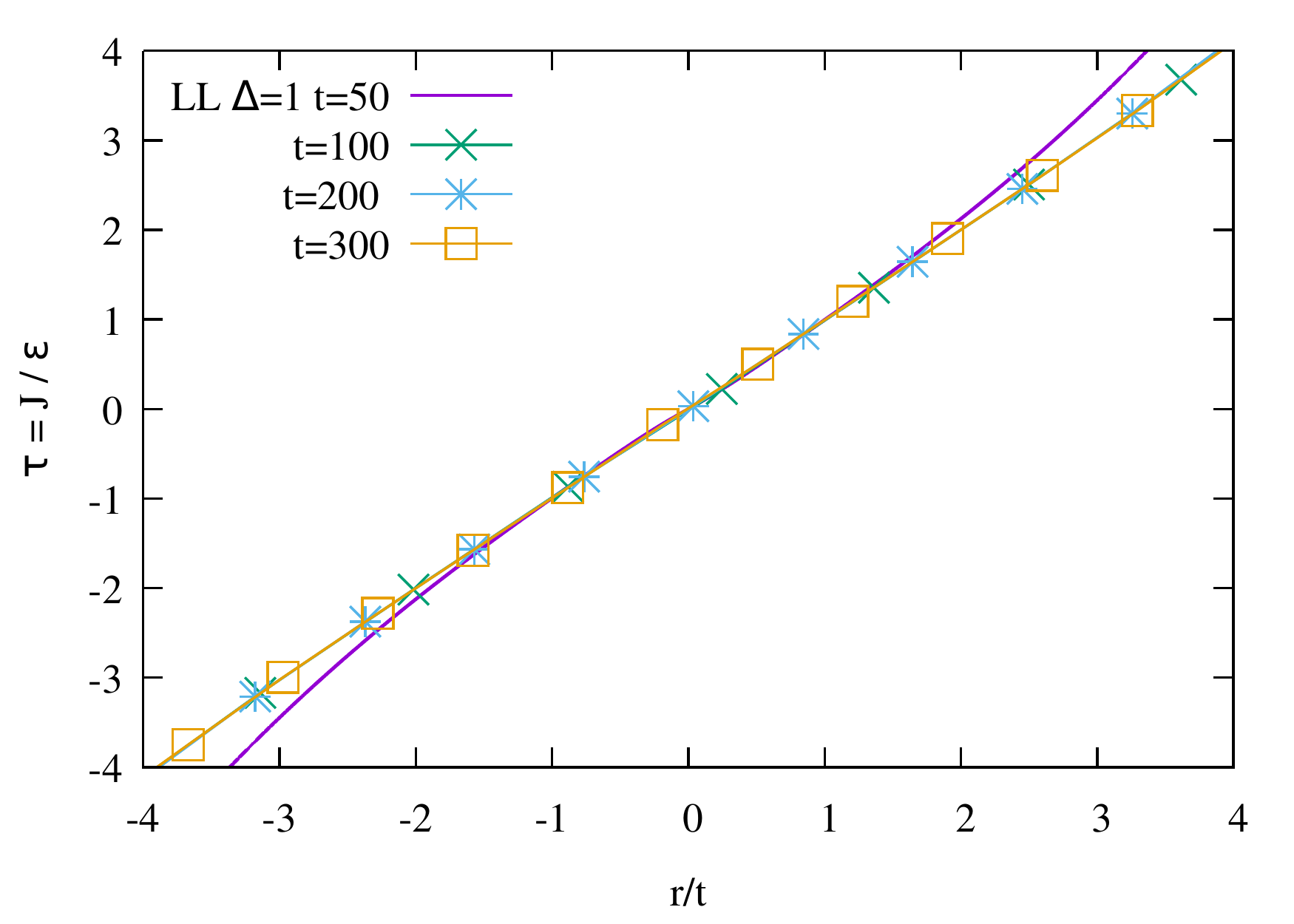}
\caption{LL torsion $\tau$ plotted as a function of $r/t$. The good
  data collapse shows that, at long times, one simply has $\tau=r/t$.
}
 \label{fig:torsion}
\end{figure}

As noted by Lakshmanan {\it et al.}~\cite{lakshmanan_1976}, it is
useful to interpret $r\to\vec M(r,t) =2\vec \Omega(r,t)$ (at fixed
$t$) as the tangent vector of a curve $\mathcal{C}$ parametrized by
$r$.  The LL energy density is then simply related to the curvature
$\kappa$ of $\mathcal{C}$:
\begin{equation}
\kappa=|| \vec M_r ||=\sqrt{8\epsilon}. 
\end{equation}
Another important geometrical property is the torsion $\tau$ of
$\mathcal{C}$ ($\tau$ should not to be confused with the rescaled time
defined in Sec.~\ref{ssec:aniLL}). Using the LL variables it reads
\begin{equation}
\tau=\kappa^{-2} \vec M \cdot 
\left( \vec M_r \wedge  \vec M_{rr}\right)=
\epsilon^{-1}\vec \Omega \cdot 
\left( \vec \Omega_r \wedge  \vec \Omega_{rr}\right).
\end{equation}
This quantity is plotted in Fig.~\ref{fig:torsion} for the LL solution
we are interested in at $\Delta=1$.  The energy current $J=\vec \Omega
\cdot \left( \vec \Omega_r \wedge \vec \Omega_{rr}\right)$ is also
plotted in Fig.~\ref{fig:J}.  The torsion data show that, at
sufficiently long times, the LL solution is well descried by a
remarkably simple relation $\tau \simeq r/t$. As discussed below, this
is reminiscent of some particular solutions of the LL problem.

The equation of motion for the magnetization is
Eq.~\eqref{eq:LL_Delta=1}, which is equivalent to $2\vec M_{t}=\vec
M\wedge\vec M_{rr}$, or to the more conventional $\vec M_{\tilde
  t}=\vec M\wedge\vec M_{rr}$ with $2\tilde t=t$.  The induced
dynamics for the curve $\mathcal{C}$ is the so-called binormal flow
\cite{gutierrez_formation_2003}, which describes the motion of a
vortex filament in the local induction approximation.  Some particular
attention has been devoted to special solutions where $\vec
M(r<0,t=0)=\vec e_z$ and $\vec M(r>0,t=0)=\cos(\gamma)\vec
e_z+\sin(\gamma)\vec e_x$
\cite{lakshmanan_1976,gutierrez_formation_2003,banica_stability_2008,de_la_hoz_numerical_2009,gutierrez_self-similar_2015,gamayun_domain_2019,gamayun_self-similar_2019}. This
is an singular initial condition, where the curve $\mathcal{C}$ forms
a sharp corner with angle $\gamma$ at $r=0$. It then evolves to a
smooth $\vec M(r,t)$ and $\mathcal C(t)$ for $t>0$. This problem can
be solved analytically for $\gamma<\pi$, and gives a self-similar
solution where $\vec M(r,t)$ is a function of $r/\sqrt{t}$ (see
Appendix \ref{sec:kummer} for an explicit expression of the self-similar
solutions in terms of Kummer confluent hypergeometric functions). It
is characterized by a spatially uniform curvature
$\kappa=E/\sqrt{\tilde t}=E/\sqrt{t/2}$ and linear torsion
$\tau=r/(2\tilde t)=r/t$ \cite{lakshmanan_1976}.  The constant $E$ is
related to the corner angle through the relation $e^{-\pi
  E^2/2}=\cos(\gamma/2)$ \cite{gutierrez_formation_2003}.  Since $E$
diverges when $\gamma$ approaches $\pi$ -- which is precisely the
situation we are interested in -- the self-similar solution breaks
down in that limit. It is thus natural to trace back the appearance of
a logarithmic correction to diffusion at $\gamma=\pi$ to this
singularity in $E$ \cite{gamayun_domain_2019}.
 
There is nevertheless a similarity between the self-similar solutions
at $\gamma<\pi$ and our case (smooth initial condition and asymptotic
angle $\gamma=\pi$): the energy density is $\sim {\rm cst}/t$ for the
self-similar solutions (at any $r/t$), while we showed that in our
case that it is $\sim \ln{t}/t$ for $r\ll t$.  But the torsion $\tau$
(Fig.~\ref{fig:torsion}) shows a more striking analogy, since we have
$\tau=r/t$ in both cases, in the regime $r\sim \sqrt{t}$ as well as
for $r/t\sim \mathcal{O}(1)$.

\begin{figure}[h]\center\includegraphics[width=0.6\linewidth]{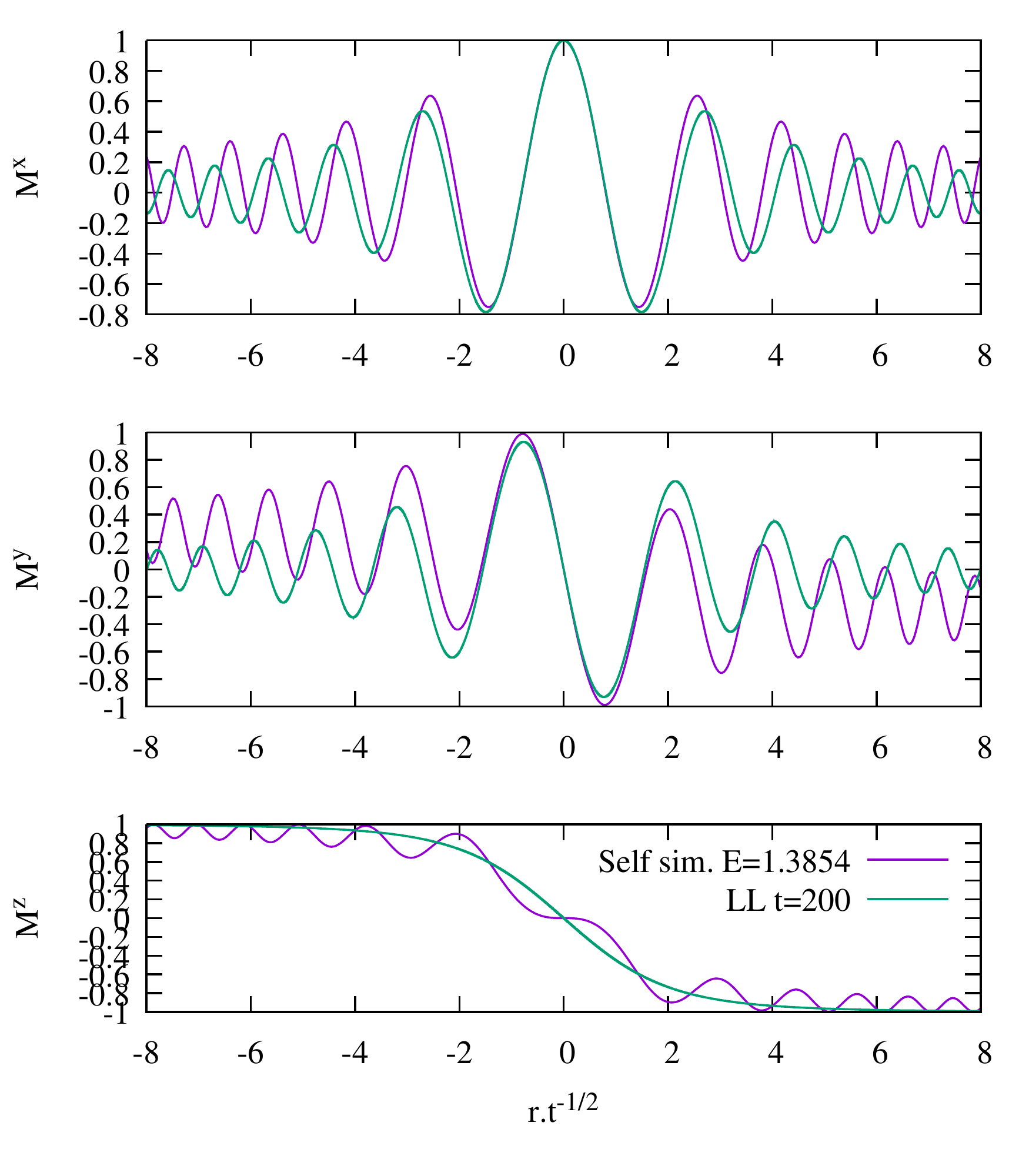}
  \caption{Components of the magnetization $\vec M(r,t)$ in the
    isotropic LL problem at $t=200$ (green lines) compared with a
    self-similar solution (see text) parametrized by $E$ (purple
    lines). The value of $E$ is chosen so that both profiles have the
    same energy density at $r=0$ (and $t=200$).  For $t=200$ we have
    $\epsilon(r=0,t)\cdot t\simeq 0.4798$ (as can be seen in the upper
    panel of Fig.~\ref{fig:FigEN_D=1}), and $E=\sqrt{ 4\epsilon\cdot
      t}\simeq 1.3854$.  }\label{fig:SS}
\end{figure}

Fig.~\ref{fig:SS} shows a comparison between the smooth domain wall
problem at $t=200$ and the self-similar solution with the same energy
density at $r=0$ (or same curvature $\kappa$). This is achieved by
taking the parameter $E$ of the self-similar solution to be
$E^2=4t\cdot\epsilon(r=0,t)$.  The two states agree by construction
when $r/\sqrt{t}$ is small, but differ otherwise. In particular,
oscillations in the $z$ component are present in the self-similar
solution, but absent from the LL we are interested in. We can also, by
inspecting the behavior of the $M^y$ component, see that the asymptotic
direction of the magnetization (when $r\to \pm \infty$) is not $\vec
M=\pm \vec e_z$ in the self-similar profile. This due to the fact
that $E$ is finite and therefore $\gamma<\pi$. Repeating this
comparison for much larger $t$ would give a larger effective $E\sim
\ln(t)$, an angle $\gamma$ closer to $\pi$, and a larger range of
$r/\sqrt{t}$ where the the two solutions are close to each other.  We
thus find that one can describe the LL magnetization profile in the
regime where $r/\sqrt{t}$ is of order one (but $r\ll t$) by
self-similar solutions with an effective time-dependent parameter
$E(t)\sim \ln(t)$.  This can however not describe the system in the
region where $r/t$ is finite, where the energy density is no longer
constant.

\subsection{Nonlinear Schrödinger equation}
\label{ssec:LL_NLS}

The isotropic LL equation is well known to be equivalent to the
Nonlinear Schrödinger (NLS)
equation~\cite{hasimoto_soliton_1972,lakshmanan_continuum_1977,zakharov_equivalence_1979}.
The mapping, also known as the Hasimoto transform, can be summarized
as follows (more details in Appendix \ref{sec:LL2NLS}).  As
explained above, we start from a solution of $2\vec M_{t}=\vec
M\wedge\vec M_{rr}$, or equivalently $\vec M_{\tilde t}=\vec
M\wedge\vec M_{rr}$ (with $2\tilde t=t$).  One then constructs the
associated curve $\mathcal{C}$ (the vortex filament) and obtains its
curvature $\kappa(r,t)$ and torsion $\tau(r,t)$.  Then one defines
the following ``filament function''  (or NLS
wave-function):
\begin{equation}\label{eq:u}
u(r,t) = \kappa(r,t)\exp{\left[i\int^r_0 \tau(r',t) dr'\right]} .
\end{equation}
On can then show that $u(r,t)$ satisfies a NLS equation:
\begin{equation}\label{eq:NLSu}
iu_{\tilde t}= 2iu_t =-u_{rr}-\frac{1}{2}u\left(|u|^2-\Lambda(t)\right),
\end{equation}
where $\Lambda(t)$ only depends on time, and is explicitly given in
terms of the curvature and the torsion close to the origin:
\begin{equation}
\Lambda(t)=
\left(2\frac{(\tau^2)_{rr}-\kappa\tau^2}{\kappa}+\kappa^2\right)_{|r=0}.
\end{equation}
Alternatively, one can define another wave-function $\psi$:
\begin{equation}
\psi(r,t)=u(r,t)\exp\left(\frac{i}{4}\int_0^t \Lambda(t')dt'\right)
\end{equation}
which is then solution of a cubic NLS equation
\begin{equation}
2i\psi_t=-\psi_{rr}-\frac{1}{2}|\psi|^2\psi.
\label{eq:NLS} 
\end{equation}

If the LL magnetization initially lies in a plane, the torsion
vanishes and the corresponding NLS wave-function can be taken to be
real at $t=0$.  We may thus write (at $t=0$) $\vec
M=(\cos(\theta(r)),\sin(\theta(r)),0)$ and the curvature is given by
$\kappa=| \theta_r |$. If we further assume that $\theta_r>0$
everywhere, the filament function is simply given by $u=\theta_r$. As
a consequence, the integral of $u$ gives the angle difference $\gamma$
between $r=-\infty$ and $r=\infty$:
\begin{equation}\label{eq:gamma}
 \gamma = \int_{-\infty}^\infty u(r,t=0) dr .
\end{equation}

With the initial condition given in Eq.~\eqref{eq:init_cond} and
$a=2$, we have $\kappa=2/\cosh(2r)$ and
\begin{equation}
 \psi(r,t=0)=u(r,t=0)=\kappa(r,t=0)=\frac{2}{\cosh(2r)}.
\label{eq:psi_t=0}
\end{equation}
This type of initial condition for the NLS equation was first analyzed
by Satsuma and Yajima~\cite{satsuma_initial_1974}, using inverse
scattering methods.  From the above remark about the integral of $u$
one can check that $\gamma = \pi$. Importantly, for this precise value
of the angle the inverse scattering approach is complicated by the
presence of  divergences~\cite{gamayun_domain_2019} and we are not
aware of an exact analytical treatment of this case.
These divergences are also responsible for the logarithmically enhanced diffusion of the $z$ magnetization.
One can however treat the case
of an angle slightly below $\pi$~\cite{gamayun_domain_2019}.

\subsection{Variational solution}
\label{sec:var_NLS}

In this section one considers the focusing NLS equation
\begin{equation}\label{eq:vs1}
i \psi_t=-\frac{1}{2\, m}\psi_{rr}-g\, |\psi|^2\psi\; , 
\quad\mbox{with}\quad m>0 \quad\mbox{and}\quad g>0\;.
\end{equation}
Eq.~\eqref{eq:NLSu} is recovered by taking $m=1$ and $g=1/4$. To
facilitate comparisons with other conventions, we will however keep
$m$ and $g$ general in the calculation below.  Equation \eqref{eq:vs1} 
is associated with
the Lagrangian density
\begin{equation}\label{vs2}
\mathscr{L}=\frac{i}{2}\left(\psi^*\psi_t-\psi\psi_t^*\right) 
-\frac{1}{2 m}|\psi_r|^2 +\frac{g}{2}|\psi|^4
=-{\rm Im}(\psi^*\psi_t)
-\frac{1}{2 m}|\psi_r|^2 +\frac{g}{2}|\psi|^4
\; ,
\end{equation}
and has a soliton solution of the form
\begin{equation}\label{vs3}
\psi_{\rm sol}(r,t)=\frac{1}{\sqrt{gm}} \frac{A_0}{\cosh(A_0 r)} 
\exp\{i A_0^2 t/2 m\} \; ,
\end{equation}
where $A_0$ is an arbitrary constant.

One considers a variational ansatz parametrized by 4 time-dependent
quantities: $A(t)$, $B(t)$, $C(t)$ and $D(t)$:
\begin{equation}\label{vs4}
\psi_{\rm var}(r,t)=\frac{1}{\sqrt{gm}} \frac{B}{\cosh(A r)} 
\exp\{i (C+D r^2)\}\; .
\end{equation}
Following the variational method proposed in
Ref.~\cite{anderson_variational_1983}, we insert the form \eqref{vs4}
into the Lagrangian density \eqref{vs2}.  It yields a Lagrangian for
the quantities $A$, $B$, $C$ and $D$:
\begin{equation}\label{vs5}
\begin{split}
L&
=\int_{\mathbb{R}}{\rm d}r \left\{
-(\dot{C}+\dot{D}r^2)|\psi_{\rm var}|^2 
-\frac{1}{2m}|\psi_{\rm var}|^2\left(A^2\tanh^2(Ar)+4 D^2 r^2\right) 
+\frac{g}{2}|\psi_{\rm var}|^4\right\}\\
& = \frac{1}{3 gm}
\left(-6\, \frac{\dot{C}B^2}{A}-\frac{\pi^2}{2}\frac{\dot{D}B^2}{A^3}
-\frac{1}{m}A B^2 -\frac{\pi^2}{m}\frac{B^2D^2}{A^3}+
\frac{2}{m}\frac{B^4}{A}\right)\\
&= \frac{1}{3 gm}
\left(-6\, \dot{C}E-\frac{\pi^2}{2}\frac{\dot{D}E}{A^2}
-\frac{1}{m}A^2 E -\frac{\pi^2}{m}\frac{E D^2}{A^2}+
\frac{2}{m} A E^2\right).
\end{split}
\end{equation}
In the last expression one has written $B=\sqrt{A E}$. The Lagrange
equations of motion corresponding to the Lagrangian $L$ show that $E$
is constant (this corresponds to the conservation of the norm $\int
|\psi_{\rm var}|^2 {\rm d}r=2 E/(gm)$) and that
\begin{subequations}\label{vs6}
\begin{align}
& 6\, \dot{C} =-\frac{\pi^2}{2} \frac{\dot{D}}{A^2} -\frac{A^2}{m}
-\frac{\pi^2}{m}\frac{D^2}{A^2}+\frac{4 A E}{m}\; , \label{vs6a}
\\
& \frac{\rm d}{{\rm d}t}
\left(\frac{E}{A^2}\right) =
\frac{4}{m} \frac{D E}{A^2}\; , \label{vs6b}\\
& 0= \frac{\pi^2\dot{D}E}{A^3}-\frac{2}{m}AE
+\frac{2\pi^2}{m}\frac{D^2E}{A^3}+\frac{2}{m}E^2\; .\label{vs6c}
\end{align}
\end{subequations}
A simple check is in order here: if one takes the initial condition
$A(0)=A_0$, $E(0)=A_0$, $C(0)=0$ and $D(0)=0$ one indeed finds the
soliton solution \eqref{vs3} with $C(t)=A_0^2 t/2m$.

In the general case, one will take $A(0)=A_0$,
$C(0)=D(0)=0$ and $E(0)=\alpha^2 A_0\equiv E_0$, where $\alpha$ is a
positive constant, so that one has initially
\begin{equation}\label{init}
\psi_{\rm var}(r,0)=\alpha\, \psi_{\rm sol}(r,0) \; .
\end{equation}

It is well known that for $\alpha$ an integer greater than one, the
initial state \eqref{init} evolves into an oscillating bound state of
$\alpha$ solitons~\cite{satsuma_initial_1974}.  The above variational
ansatz cannot treat such cases. On the other hand, for $\alpha$ close
to 1 we expect this approximation to correctly describe the nonlinear
dynamics.  In terms of the LL description, $\alpha$ parametrizes a
family of planar initial magnetization profiles
\begin{equation}\label{initM}
\vec{M}(r,t=0)=\begin{pmatrix}\sin\theta(r)\\
0\\
\cos\theta(r)\end{pmatrix} \quad\mbox{with}\quad
\theta(r)=2\alpha \arccos\left(\tanh(-A_0 r)\right)\; .
\end{equation}
with an angle difference $\gamma=2\alpha\pi$ (as expected from
\eqref{eq:gamma}).

Here we will push the approximation to $\alpha=\frac{1}{2}$, which is
the case of interest since the initial condition of
Eq.~\eqref{eq:psi_t=0} corresponds to $A_0=2$ and $\alpha=1/2$.  Said
differently, we consider a case where $\psi(t=0)$ is ``half'' of a
static soliton solution.

The next section explains how to solve the equations
Eqs.~\eqref{vs6}. The comparison with the numerical results will be
done in Sec.~\ref{ssec:var_discuss}.

\subsubsection{Solution of the variational dynamics}\label{ssec:var_sol}
\begin{figure}[h]\center
\includegraphics[height=0.35\linewidth]{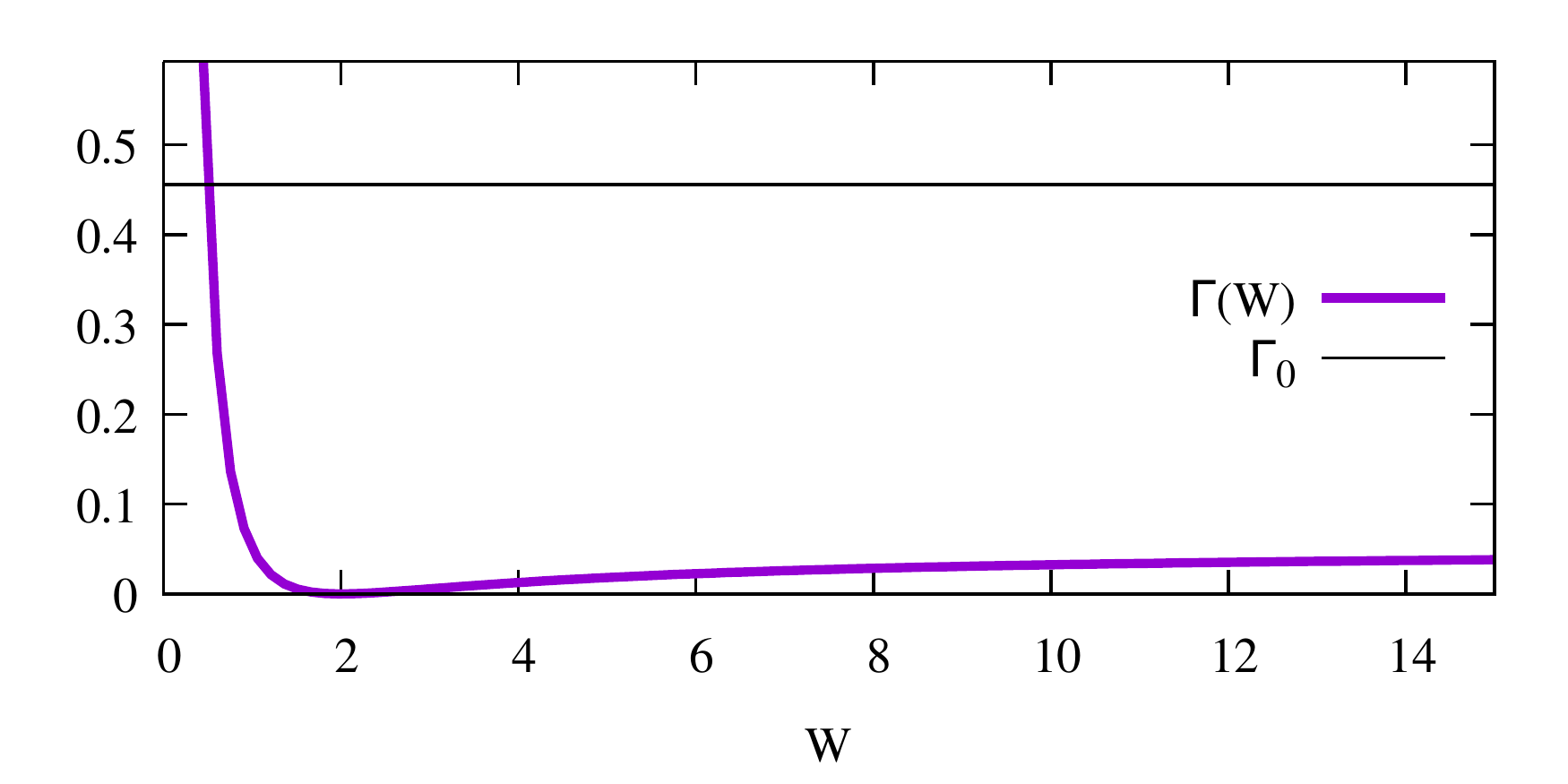}
\caption{Effective potential $\Gamma$ for the width $W(t)$ [see text
  and Eq.~\eqref{eq:vs11}] for $m=1$, $g=1/4$, $A_0=2$ and
  $\alpha=1/2$.}
 \label{fig:Gamma}
\end{figure}

Factorizing out $E=E_0$ from Eq. \eqref{vs6c} and rewriting
\eqref{vs6a} and \eqref{vs6c} one gets
\begin{subequations}\label{vs7}
\begin{align}
 -\frac{\pi^2}{2A^2}\left(\dot{D}+\frac{2D^2}{m}\right)
+\frac{A}{m}(4E_0-A)& =6\,\dot{C}\; ,\label{vs7a}\\
 \frac{\pi^2}{A^3}\left(\dot{D}+\frac{2D^2}{m}\right)
+\frac{2}{m}(E_0-A)& =0\; , \label{vs7b}
\end{align}
\end{subequations}
which shows that
\begin{equation}\label{vs8}
6\,\dot{C}=\frac{A}{m}(5 E_0-2 A)\; .
\end{equation}
Eq. \eqref{vs6a} shows that
\begin{equation}\label{eqD}
D=-m \dot{A}/(2 A)\; ,
\end{equation} 
which, when inserted
into \eqref{vs7b} yields
\begin{equation}
\frac{\rm d}{{\rm d}t}\left(-\frac{\dot{A}}{A^2}\right)+
\frac{4 A^2}{m^2 \pi^2}(E_0-A)=0\; .
\end{equation}
Defining the width $W=1/A$ of the variational solution \eqref{vs4}, this reads
\begin{equation}
\ddot{W}+\frac{4}{m^2\pi^2W^2}\left(E_0-\frac{1}{W}\right)=0\;
\end{equation}
which admits the first integral
\begin{eqnarray}\label{eq:vs11}
\frac{\dot{W}^2}{2}+\Gamma(W)&=&\Gamma_0 \\
\quad\mbox{where}\quad
\Gamma(W)&=&\frac{2}{m^2\pi^2}\left(\frac{1}{W}-\alpha^2 A_0\right)^2 \\
\quad\mbox{and}\quad
\Gamma_0&=&\frac{2 A_0^2}{m^2\pi^2}(\alpha^2-1)^2\; .
\end{eqnarray}
It is easy to see that if $\alpha<1/\sqrt{2}$ one has
$\Gamma_0>\Gamma(+\infty)$ and the motion is not bounded. Such a
situation is illustrated in Fig.~\ref{fig:Gamma}. Since the case of
interest is $\alpha=1/2$ we are in a situation of this type.

The exact dynamics indeed shows two regimes, depending on the value of
$\alpha$.  For $\alpha\leq\frac{1}{2}$ the solution spreads
monotonously, while for $\frac{3}{2}>\alpha>\frac{1}{2}$ the nonlinear
structure oscillates around a one-soliton
solution~\cite{satsuma_initial_1974,yuan_interplay_2019}.  The
variational approach thus qualitatively captures the two regimes, but
not the exact value of the transition point. Still, for $\alpha=1/2$,
the present variational calculation correctly predicts an unbounded
spreading.

To get the time evolution of $W$ one writes $\sqrt{2}\, {\rm d}t={\rm
  d}W/\sqrt{\Gamma_0-\Gamma(W)}$. In the case $\alpha=1/2$ we are
interested in, defining
\begin{equation}
w=W/W_0=A_0 W, 
\end{equation}
one finds
\begin{eqnarray}\label{dyn}
\frac{\sqrt{2}}{m\pi W_0^2}\, \int_0^t {\rm d}t &=& \nonumber
\int_1^w \frac{w \, {\rm d}w}{\sqrt{(w-1)(w+2)}} \\
&=&
\sqrt{w^2+w-2}-\frac{1}{2}\ln\left(1+2w+2\sqrt{w^2+w-2}\right)+
\frac{\ln 3}{2}\; .
\end{eqnarray}
This implies that, at large times, $W(t)\propto t$ with (additive)
logarithmic corrections.  Then \eqref{eqD} yields
\begin{equation}\label{eddt}
D(t)=\frac{m}{2}\frac{\dot{w}}{w}=
\frac{\sqrt{(w-1)(w+2)}}{\pi\sqrt{2}\, W_0^2 w^2}\; ,
\end{equation}
which implies that, at large times, $D\propto t^{-1}$ with logarithmic
corrections. For solving the variational problem one just needs to
invert Eq. \eqref{dyn} to get $w=w(t)$; then $A=A_0/w$, $D$ is given from
\eqref{eddt} and from \eqref{vs8} one gets
\begin{equation}
\frac{{\rm d}C}{{\rm d}w}=\frac{\pi}{6\sqrt{2}} \, 
\frac{1}{\sqrt{(w-1)(w+2)}}\left(\frac{5}{4}-\frac{2}{w}\right)\; ,
\end{equation}
which makes it possible to express $C$ as a function of $w$.  The
expression is cumbersome, but at large $t$ it is of the form ${\rm
  cst}+\ln(t)$.

\subsubsection{Comparison with numerics}\label{ssec:var_discuss}

As we explain now, the variational treatment discussed above
reproduces a number of important features of the numerical solution.
The modulus of the NLS wave-function is the curvature $\kappa$ in the
LL problem: $|\psi|=\kappa$. In the variational approach we have
$|\psi_{\rm
  var}|=\sqrt{\frac{E_0}{gmW(t)}}\left\{\cosh\left(r/W(t)\right)\right\}^{-1}$.
Since the LL energy density is $\epsilon=\frac{1}{8}\kappa^2$ we can
express the energy density using $W(t)$: $\epsilon^{-1}\simeq 4 W(t)
\cosh\left(r/W(t)\right)^2$ (we have used $E_0=1/2$).  Since $W(t)$
grows linearly with time, the variational treatment correctly finds
that $\epsilon\cdot t$ is a function of $r/t$.

For $r\ll t$, $|\psi_{\rm var}|$ and the curvature $\kappa$ are
approximately independent or $r$, and equal to $\kappa
=\sqrt{\frac{E_0}{gmW(t)}}$, which goes to zero as $1/\sqrt{t}$.  On
the other hand, when the curvature is constant in space we may
estimate the length scale over which the magnetization goes from the
vicinity of $+\vec e_z$ to the vicinity of $-\vec e_z$ by
$\kappa^{-1}$.\footnote{This amounts to approximate
this part of the curve by a half circle.}
Here this length scale is proportional to
$\sqrt{t}$. In other words the variational approximation predicts a
diffusive expansion of the $M^z$ profile.  As discussed previously
this is correct up to logarithmic correction.

\begin{figure}[h]\center
\includegraphics[height=0.35\linewidth]{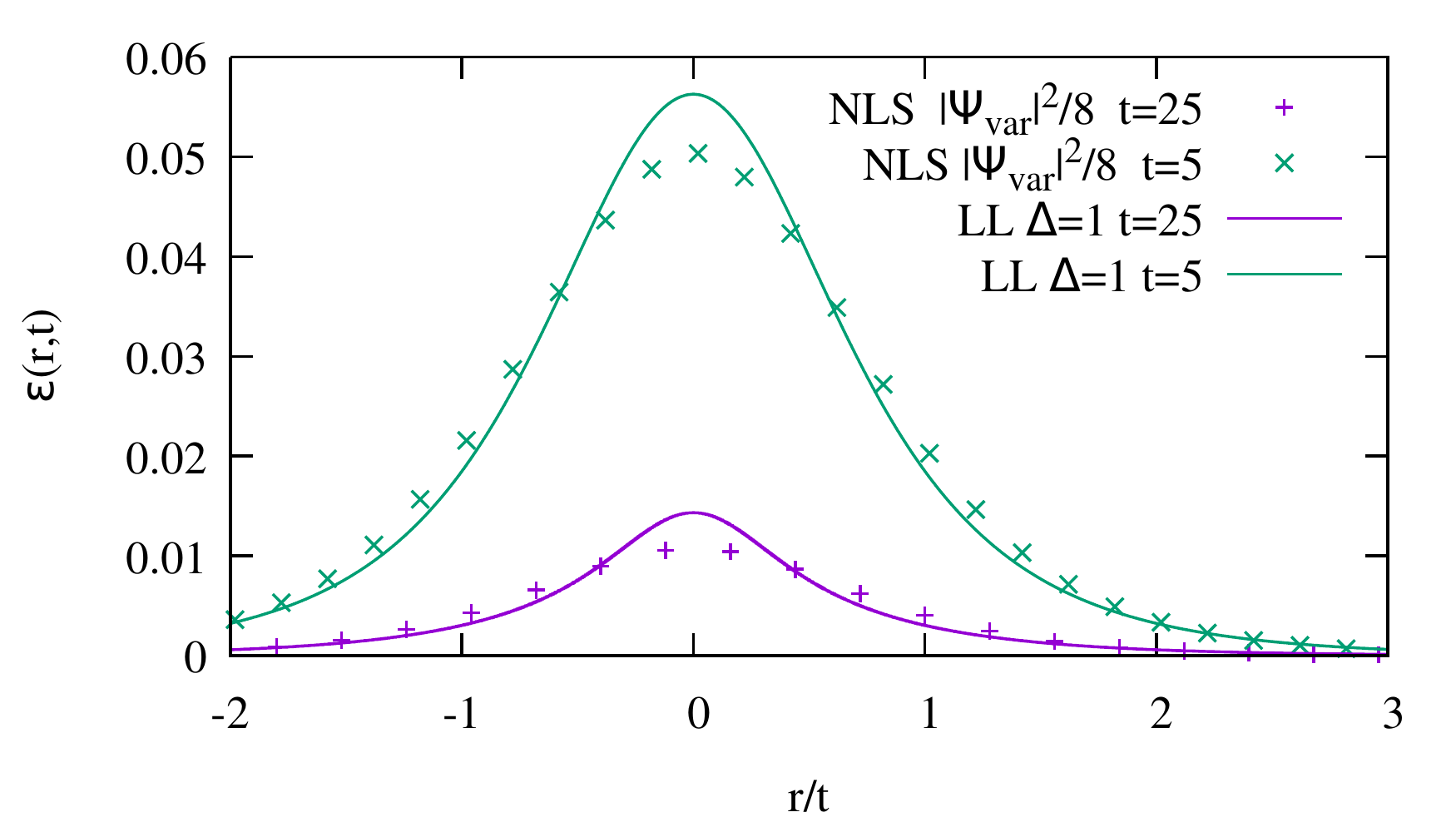}
\caption{Energy density $\epsilon$ in the isotropic LL model compared
  with $|\psi_{\rm var}|^2/8$ from the variational solution of the NLS
  equation [Eq.~\eqref{vs4}].  }
 \label{fig:var_nls}
\end{figure}

Going back to the energy density away from the origin, the variational
approximation gives an exponentially decay at large $r/t$, which is in
agreement with the exact result discussed in
Sec.~\ref{ssec:energy_density}.  Using Eq.~\eqref{dyn} we find
 \begin{equation}\label{dynaa}
W(t)\simeq \frac{2\sqrt{1-2\alpha^2}}{m\pi W_0}\, t\; ,
\quad\mbox{when}\; t\to +\infty\; .
\end{equation}
For $\alpha=1/2$, $W_0=1/2$ and $m=1$ this gives $W(t)\sim t\cdot
2\sqrt{2} / \pi$.  For large $r/t$ we get $\epsilon\cdot t\sim
\exp(-c\;r/t)$ with $c=\pi/\sqrt{2}\simeq 2.22$, to be compared to
$\simeq 1.6$ obtained from the fit in Fig.~\ref{fig:FigENlog_D=1}.
This technically simple and physically intuitive approach thus
reproduces the scaling of the energy profile at large distance.  The
modulus squared of the NLS wave-function, $|\psi_{\rm var}|^2/8$, is
compared with the energy density of the (isotropic) LL problem in
Fig.~\ref{fig:var_nls}.  By construction the two curves coincide at
$t=0$ (not shown), and they remain close to each other at short times.
The variational solution captures qualitatively the spread of the
energy at long times. It does however not capture the logarithmic term
observed in Fig.~\ref{fig:Energy_density_r=0}, hence the fact that the
LL energy density at $r=0$ decays more slowly with time ($\sim \ln(t)/t$) than the variational
estimate ($\sim 1/t$).
  
It is also useful to consider the NLS current $J=\frac{1}{m}{\rm
  Im}(\psi^*\psi_x)$. With the variational Ansatz it becomes
\begin{equation}\label{rj}
J_{\rm var}(r,t)=\frac{2}{m}D\, r |\psi_{\rm var}|^2\; .
\end{equation}
Eqs. \eqref{dynaa}, \eqref{eqD} and the above show that
$D(t)=\frac{m}{2}{\rm d}\ln \left(w\right)/{\rm d}t \simeq m/(2t)$ at
large $t$. These asymptotic estimates combined with \eqref{rj} yield
\begin{equation}\label{rhoJ0}
J_{\rm var}/|\psi_{\rm var}|^2\simeq r/t,
\end{equation}
independently of the value of $\alpha$ and $W_0$.  Since this ratio is
also the space derivative $\theta_r$ of the complex argument of
$\psi$, it gives, according to Eq.~\eqref{eq:u}, the LL torsion. We
thus recover the relation $\tau\simeq r/t$ obeyed by the solution of
the LL problem (Fig.~\ref{fig:torsion}).

\section{Conclusion}

In this work we argued that the dynamics of the domain wall problem
for a spin-$\frac{1}{2}$ XXZ chain close to the isotropic point
behaves semiclassicaly at long time, and can asymptotically be
described in the framework of an LL equation.  This picture is
supported by the systematic numerical comparison we made between the
quantum and the classical problems, including an analysis of the
in-plane component of the magnetization, the energy density or the
energy current.

We identified the different scaling regimes for all these quantities,
from the shortest length scale with $r\sim\mathcal{O}(1)$ to larger
distances, from $r\sim\mathcal{O}(\sqrt{t})$ to $r\sim\mathcal{O}(t)$.
The latter regime, that we call ballistic, is present in the LL problem
as well as in the quantum model, in the easy-plane, isotropic and
easy-axis cases. This regime, where the magnetization is almost
pointing along the $z$ axis, can be captured by a perturbative
expansion or, at $\Delta=1$, by a simple variational calculation based
on the isotropic LL to NLS mapping.  The latter approximation predicts
a diffusive expansion of the $z$ magnetization profile but misses the
observed logarithmic correction (in the energy density, or in the
width of the $z$ profile).  Still at $\Delta=1$, we showed that the LL
magnetization in the diffusive regime can be described by a
self-similar solution with an effective time-dependent curvature which
incorporates the logarithmic enhancement.  But, although a logarithmic
divergence was identified in the inverse scattering data of the
isotropic LL problem~\cite{gamayun_domain_2019}, it seems
that a simple and/or intuitive explanation for the appearance of the
logarithmic terms at $\Delta=1$ is still missing, as well as an
analytical expression for the asymptotic magnetization profile.

There are several possible future directions for developing the line
of research exposed in the present work. The anisotropic LL problem has a
great richness and can give rise to several other types of nonlinear
phenomena, like dispersive shock waves or rarefaction wave, see for
instance Ref. \cite{ivanov_solution_2017}. We could also mention some
fascinating polygonal structures that can emerge in the isotropic case
for particular initial conditions~\cite{hoz_vortex_2014}.  It would be
interesting to investigate if some other initial conditions in the
quantum spin chain could realize these phenomena.

Exploring the quantum corrections to the classical behavior in the
quantum chain also seems a promising direction for future studies, and
could take the form of some $1/S$ expansion.  As for the entanglement
entropy, while the heuristic argument given in
Sec.~\ref{ssec:entanglement} in favor of $S_{\rm vN}\sim
\frac{1}{2}\ln t$ is simple and appealing, this prediction does not
fully agree with the present numerics. The entanglement growth is
certainly one of the interesting open questions in the domain wall
problem at $\Delta=1$.

\section*{Acknowledgements}
We thank V. Banica, A.M. Kamchatnov, K. Mallick, P. Krapivsky, J.-M. Stéphan and J. de
Nardis for discussions about this problem.  We are particularly
grateful to K. Mallick and P. Krapivsky for sharing their unpublished
results~\cite{mallick_2017}.

\appendix

\section{From isotropic LL to NLS}\label{sec:LL2NLS}

\newcommand\g{\mathfrak{g}} 
\newcommand\M{\bold{M}} 
\newcommand\J{\bold{J}}
\newcommand\LL{\mathbb{L}} 
\newcommand\MM{\mathbb{M}} 

For completeness we provide here a detailed and self-contained
derivation of the mapping from the isotropic LL equation to NLS~\cite{hasimoto_soliton_1972,lakshmanan_continuum_1977,zakharov_equivalence_1979}.  The
starting point is the isotropic LL equation:
\begin{equation}\label{eq:2mMt}
  2m \vec M_t = \vec M\wedge \vec M_{rr}.
\end{equation}
with $\vec M^2(r,t)=1$.  The parameter $m$ was introduced to simplify
the comparison with other possible conventions. To match the
convention used in Sec.~\ref{ssec:filament} one has to take $m=1$.

To the 3-vector $\vec M=(M^1,M^2,M^3)$ we associate a $2\times2$ matrix
\begin{equation}\label{eq:M123}
 \M=M^1\sigma_1 +M^2\sigma_2+M^3\sigma_3,
\end{equation}
where $\sigma_1$, $\sigma_2$ and $\sigma_3$ are the Pauli matrices.
The magnetization $\vec M$ can be obtained by a rotation the $(0,0,1)$
vector. In terms of the associated 2 by 2 matrix, we can thus write
\begin{equation}\label{eq:Mgsg}
 \M(r,t)=\g^{-1} \;\sigma_3 \;\g
\end{equation}
where $\g(r,t) \in SU(2)$ implements the rotation. The space
derivative of $\M$ is then
\begin{equation}\label{eq:Mr1}
 \M_r = \g^{-1} \left[\sigma_3, \LL \right]  \g
\end{equation}
where we have defined
\begin{equation}\label{eq:Ldef}
\LL(r,t)=\g_r \; \g^{-1}.
\end{equation}
In a similar way the time derivative of the magnetization matrix reads
\begin{equation}\label{eq:Mt}
\M_t =  \g^{-1} \left[\sigma_3, \MM \right]  \g
\end{equation}
where we have defined
\begin{equation}\label{eq:Mdef}
\MM(r,t)=\g_t \; \g^{-1}.
\end{equation}

There is gauge freedom in the choice of the rotation since
$\g'=\g\exp(i\Theta \sigma_3)$ with any angle $\Theta(r,t)$ would also
be a valid choice in \eqref{eq:Mgsg}. In the following we chose a
gauge such that the matrix $\LL$ defined above has zeros on its
diagonal:
\begin{equation}\label{eq:Ldef2}
\LL=\left( \begin{matrix} 0 & -\sqrt{gm}\;\bar u(r,t) \\ \sqrt{gm}\;u(r,t) & 0\end{matrix} \right).
\end{equation}
The parameter $g$ is arbitrary here and is simply introduced here to
facilitate the comparison with other conventions found in the
literature concerning the NLS equation. Note that the convention used
in Sec. \ref{ssec:filament} amounts to take $g=1/4$ and $m=1$.

With the above gauge choice $\LL$ anti-commutes with $\sigma_3$ and
\begin{equation}\label{eq:s3L}
\left[\sigma_3,\LL \right]=-2\sqrt{gm}\;
\left(u \sigma^+ +\bar u \sigma^-\right)
\end{equation}
with $\sigma^+=\left(\begin{matrix} 0 & 0 \\ 1 & 0\end{matrix}\right)$
and $\sigma^+=\left(\begin{matrix} 0 & 1 \\ 0 &
    0 \end{matrix}\right)$.  It is interesting to compute $\vec
M_r^2$, since it proportional to the energy energy density in LL.
From \eqref{eq:Mr1} we have $\vec M_r=-\det\left( \left[\sigma_3,\LL
  \right] \right)$ and thus, from \eqref{eq:s3L},
\begin{equation}\label{eq:Mr2-4gmu2}
 \vec M_r^2=4gm |u|^2.
\end{equation}
As a check, if we use the convention of Sec.~\ref{ssec:LL_NLS},
i.e. $g=1/4$ and $m=1$, the equation above gives a curvature
$\kappa=||\vec M_r||=|u|$. This is consistent with Eq.~\eqref{eq:u}.

Combining \eqref{eq:Mr1} and \eqref{eq:s3L} we get
\begin{equation}\label{eq:Mr2}
 \M_r = -2 \sqrt{gm}\;\g^{-1} \left( u \sigma^+ +\bar u \sigma^-\right)  \g.
\end{equation}
Now we come back to the LL equation of motion. It can be written in
terms of the magnetization current
\begin{equation}\label{eq:J}
\vec J = \frac{1}{2m} \vec M \wedge \vec M_{r}, 
\end{equation}
such that Eq. \eqref{eq:2mMt} becomes $\vec M_t = \vec J_r$. From
\eqref{eq:J} we get the matrix associated to $\vec J$ as a commutator:
\begin{eqnarray}
\J &=& \frac{1}{4im}\left[\M,\M_r \right].
\end{eqnarray}
Using Eqs. \eqref{eq:Mgsg} and \eqref{eq:Mr2} this is expressed as
\begin{eqnarray}
\J&=&\frac{-2\sqrt{gm}}{4im} \g^{-1} 
\left[ \sigma_3, u \sigma^+ +\bar u \sigma^-\right] \g \\
&=& \frac{1}{im} \g^{-1} \LL  \g \;= \frac{1 }{im} \g^{-1} \g_r
\end{eqnarray}
As for the current gradient, it reads
\begin{equation}
 \J_r = \frac{1}{im} \g^{-1} \LL_r  \g.
\end{equation}
Using Eq. \eqref{eq:Mt} and the equation above, we can write the
equation of motion $\M_t=\J_r$ as
\begin{eqnarray}
\left[\sigma_3,\MM\right]=  \frac{1}{im} \LL_r.
\end{eqnarray}
Using the explicit form of $\LL$ \eqref{eq:Ldef2}, the equation above
implies that $\MM$ must have the following form:
\begin{equation}
 \MM = i\left(\begin{matrix}
 \alpha & \frac{1}{2}\sqrt{ \frac{g}{m} }\bar u_r \\
 \frac{1}{2}\sqrt{ \frac{g}{m} }u_r & -\alpha
 \end{matrix} \right),
\end{equation}
where, so far, $\alpha$ is unknown.  The fact that $\MM$ must be
traceless can be obtained from \eqref{eq:Ldef}, since it implies that
${\rm Tr}\MM=\frac{d}{dt}{\rm det}\left(\g\right)$. The latter
derivative vanishes since ${\rm det}\left(\g\right)=1$.  At this point
we will consider two different expressions for $\g_{rt}$.  The first
one can be obtained by taking $\g_r$ from \eqref{eq:Ldef} and then
taking the derivative with respect to time:
\begin{equation}
 \g_{rt} = \frac{d}{dt} \left( \LL \g \right)
 = \LL_t \g + \LL \g_t \;= \LL_t \g + \LL \MM \g \label{eq:grt1}
\end{equation}
where, in the last expression, we have written $\g_t$ using
\eqref{eq:Mdef}. Now we repeat the calculation of $\g_{rt}$, doing the
derivations in reverse order:
\begin{equation}
 \g_{rt} = \frac{d}{dr} \left( \MM \g \right)
= \MM_r \g + \MM \g_r \;= \MM_r \g + \MM \LL \g \label{eq:grt2}
\end{equation}
Comparing \eqref{eq:grt1} and \eqref{eq:grt2} we get the Lax equation:
\begin{equation}
 \LL_t-\MM_r+\left[ \LL,\MM\right]=0.
\end{equation}
The upper left element of the matrix equation above is
\begin{equation}
 -i\alpha_r-\frac{ig}{2}(u\bar u)_r=0.
\end{equation}
It means that $\alpha=-\frac{g}{2} |u|^2 +A(t)$ where $A$ is some
integration constant, independent of $r$.  We then write down the
lower left element of the Lax equation:
\begin{equation}
\sqrt{gm} u_t-\frac{i}{2}\sqrt{ \frac{g}{m} } u_{rr} +2iu\alpha \sqrt{gm} =0.
\end{equation}
Replacing $\alpha$ by its expression found above we get:
\begin{equation}
\sqrt{gm} u_t-\frac{i}{2}\sqrt{ \frac{g}{m} } u_{rr} 
+2iu(-\frac{g}{2} |u|^2 +A(t))\sqrt{gm} =0
\end{equation}
and after some rearrangement:
\begin{equation}\label{eq:NLSu2}
i u_t = -\frac{1}{2m} u_{rr} -g u \left(|u|^2 -\Lambda(t)   \right)
\end{equation}
with $\Lambda(t)=2A(t)/g$. The equation above is the same NLS equation
as \eqref{eq:NLSu} if we set $m=1$ and $g=1/4$.

To finish this short review of the mapping, it is interesting to
interpret geometrically the (space- and time-dependent) $SO(3)$
rotation induced by $\g$, which rotates the vector $(0,0,1)$ to the
magnetization direction $\vec M$.  To this end one defines the
orthonormal basis $\left\{\vec e_1, \vec e_2, \vec M \right\}$ which
is the image of the initial basis under the rotation. We then consider
the $2\times2$ matrices associated to $\vec e_1$ and $\vec e_2$:
\begin{equation}
 \bold{e}_1= \g^{-1} \sigma_1 \g \;\;\;{\rm and}\;\;\;
 \bold{e}_2= \g^{-1} \sigma_2 \g.
\end{equation}
Taking the space derivative of the equations above (as was done in
\eqref{eq:Mr1}) we get:
\begin{equation}
 \bold{e}_{1 r}=\g^{-1} \left[\sigma_1, \LL \right]  \g \;\;\;{\rm and}\;\;\;
 \bold{e}_{2 r}=\g^{-1} \left[\sigma_2, \LL \right]  \g.
\end{equation}
Using the explicit form \eqref{eq:Ldef2} of $\LL$ (Eq.~\ref{eq:Ldef2}) we find
\begin{equation}
\begin{split}
 \bold{e}_{1 r}&=2\sqrt{gm}\; {\rm Re}(u) \sigma_3 , \\
 \bold{e}_{2 r}&=2\sqrt{gm}\; {\rm Im}(u) \sigma_3.
\end{split}
\end{equation}
Going back to 3-vectors we can thus write:
\begin{equation}
 \left(\begin{matrix}
 \vec e_1 \\ \vec e_2 \\ \vec M  
 \end{matrix}\right)_r
=2\sqrt{gm}\;\left(\begin{matrix}
        0 & 0 & {\rm Re}(u) \\
        0 & 0 & {\rm Im}(u)\\
        -{\rm Re}(u)  & -{\rm Im}(u) & 0
       \end{matrix}
\right)\left(\begin{matrix}
 \vec e_1 \\ \vec e_2 \\ \vec M  
 \end{matrix}\right)
\end{equation}
where, to obtain the last line, we have used the fact that the
$3\times3$ matrix above, describing the space derivative of the
rotation which connects the initial basis to the local one, must be
antisymmetric.  We see in particular that the gauge choice made in
\eqref{eq:Ldef2} does not correspond to the local Serret-Frenet frame
of the curve, but to its Bishop frame \cite{bishop_there_1975}.

\section{From NLS to isotropic LL}
\label{sec:NLS2LL}

In this section we recall how, starting from a solution of the NLS
equation, one can construct the associated solution of the isotropic
LL equation.  We start from a solution $u(r,t)$ of Eq.~\eqref{eq:NLSu}
and we wish to construct the associated LL magnetization $\vec M$.
The starting point is Eq.~\eqref{eq:Ldef}, that we rewrite under the form
\begin{equation}\label{eq:gr}
 \g_r=\left( \begin{matrix} 0 & -\sqrt{gm}\;\bar u \\ \sqrt{gm}\;u & 0
\end{matrix} \right) \g.
\end{equation}
We parametrize $\g$ using two complex numbers $\mu$ and $\nu$ satisfying 
$|\mu|^2+|\nu|^2=1$ as follows:
\begin{equation}
\g=\left( \begin{matrix} 
 \mu & -\bar \nu \\
 \nu & \bar \mu
\end{matrix} \right).
\end{equation}
Looking at the first column of Eq.~\eqref{eq:gr}, we get the following
equations for $\mu$ and $\nu$ and their space derivatives:\footnote{Note that $(\mu,\nu)$ may be viewed a quantum state for a
  spin-$\frac{1}{2}$: $|\phi\rangle = \left(\begin{matrix} \mu \\
      \nu\end{matrix} \right)$. If we interpret $r$ as a fictitious
  time, the equations above describe the motion of a spin in some
  ``time''-dependent magnetic field $\vec B(r)$.  Indeed we have
  $i\frac{d}{dr}|\phi\rangle=H|\phi\rangle$ with an Hamiltonian
  $H=B^1\sigma_1+B^2\sigma_2=\sqrt{gm}\left[ {\rm Re}(u)\sigma_2-{\rm
      Im}(u)\sigma_1\right]$, that is $\vec B=\sqrt{gm} ({\rm
    Re}(u),-{\rm Im}(u),0)$.  }
\begin{eqnarray}
 \mu_r &=& - \sqrt{gm}\; \bar u \;\nu \label{eq:mur} \\
 \nu_r &=&  \sqrt{gm}\; u \;\mu.
\end{eqnarray}
After eliminating $\nu$ from these equations we obtain a linear
differential equation for $\mu$:
\begin{equation}\label{eq:mueq}
 \mu_{rr}-\mu_r \frac{\bar u_r}{\bar u	} +gm |u|^2 \mu=0.
\end{equation}
Note that the above derivation has the advantage of giving directly a
{\em linear} differential equation for $\mu$.  We are interested in
situations where the magnetization points in the $x$ direction at
$r=0$.  It is then convenient to adopt the following correspondence
between the directions $1,2,3$ of Eq.~\eqref{eq:M123} and the
direction $x,y,z$: $x\to3$, $-y\to1$ and $-z\to2$. In this way, $\g(r=0)$
is the identity and the initial condition for $\eqref{eq:mueq}$ is
$\mu(r=0)=1$ (and $\nu(r=0)=0$).  So, using Eqs.~\eqref{eq:M123} and
\eqref{eq:Mgsg} we can relate $\mu$ and $\nu$ to the magnetization:
\begin{subequations}\label{eq:munu2M}\begin{eqnarray}
 -M^1=M^y&=&2\;{\rm Re}\left(\mu \nu \right) \\
 -M^2=M^z&=&2\;{\rm Im}\left(\mu \nu \right) \\
 M^3=M^x&=&|\mu|^2 - |\nu|^2=2|\mu|^2-1.
\end{eqnarray}
\end{subequations}

\section{Magnetization for the self-similar solutions of the isotropic LL equation}
\label{sec:kummer}

We provide here a detailed solution of Eq.~\eqref{eq:mueq}, making it
possible to obtain some explicit expression for the self-similar
solutions of the isotropic LL equation. The final result
[Eqs.~\eqref{eq:mu_F1} and \eqref{eq:nu_F1}] is expressed in terms of
some confluent hypergeometric functions (Kummer function
${_1}F_1$). Some of these results have been found previously in
Refs.~\cite{gutierrez_formation_2003,gamayun_self-similar_2019}.

We start with the following filament  function, parametrized by $E$:
\begin{equation}\label{eq:uss}
 u(r,t)=\frac{E}{2\sqrt{t}} \exp\left(\frac{ir^2}{4t}\right).
\end{equation}
It is a solution of
\begin{equation}
 iu_t = - u_{rr} -g u\left( |u|^2-\frac{E^2}{4t} \right),
\end{equation}
It corresponds to $m=\frac{1}{2}$ and $\Lambda(t)=\frac{E^2}{4t}$ in
Eq.~\eqref{eq:NLSu2}.  We then set $g=2$.  Since $\bar u_r /\bar
u=-ir/(2t)$, when we plug \eqref{eq:uss} into \eqref{eq:mueq} we
obtain \cite{gutierrez_self-similar_2015}:
\begin{equation}
\mu_{rr} +\frac{ir}{2t}  \mu_r +\frac{E^2}{4t} \mu=0.
\end{equation}
We then replace the variable $r$ by $z=-ir^2/(4t)$. After a few manipulations
we arrive at
\begin{equation}
 \mu_{zz} +\mu_z \left(\frac{1}{2}-z\right) + \frac{iE^2}{4}\mu=0.
\end{equation}
With the initial condition $\mu(0)=1$, the solution of this equation
is the Kummer (confluent hypergeometric) function
\begin{equation}\label{eq:mu_F1}
 \mu(r,t)={_1}F_1(-\frac{iE^2}{4} , \frac{1}{2},-\frac{ir^2}{4t}). 
\end{equation}
We can, using Eq.~\eqref{eq:mur}, get the other matrix element of
$\g$, which satisfies $\nu(r=0,t)=0$:
\begin{equation}
  \nu(r,t)=-\frac{\mu_r}{\bar u}
  = \frac{2\sqrt{t}}{E}e^{\frac{ir^2}{4t}} \mu_r.
\end{equation}
And since $\frac{d}{dz}\;{_1}F_1(p,q,z)=
\frac{p}{q}\;{_1}F_1(p+1,q+1,z)$ we
find~\cite{gamayun_self-similar_2019}:
\begin{eqnarray}\label{eq:nu_F1}
  \nu(r,t)&=& \frac{2\sqrt{t}}{E} 
\left( \frac{-iE^2}{2}\right)\left(\frac{-ir}{2t}\right)  
e^{\frac{ir^2}{4t}}  {_1}F_1(-\frac{iE^2}{4} 
+1, \frac{3}{2},-\frac{ir^2}{4t}) \nonumber \\
  &=& -\frac{ar}{2\sqrt{t}}
  e^{\frac{ir^2}{4t}}  {_1}F_1(-\frac{iE^2}{4} +1, \frac{3}{2},-\frac{ir^2}{4t})
\end{eqnarray}
These expressions for $\mu$ and $\nu$ can be used to obtain the
components of $\vec M$, using Eqs.~\eqref{eq:munu2M}. A comparison
between such a self-similar solution and the solution of the LL domain
wall problem with a smooth initial condition is shown in
Fig.~\ref{fig:SS} and is discussed at the end of
Sec.~\ref{ssec:filament}.

As a consistency check, one can verify that the formula above yields a
curvature $||\vec M_r||=\kappa=E/\sqrt{t}$. This is consistent with
Eq.~\eqref{eq:uss} and Eq.~\eqref{eq:Mr2-4gmu2} with $g=2$ and
$m=1/2$.  One can also check that the torsion $\tau=\kappa^{-2} \vec M
\cdot \left( \vec M_r \wedge \vec M_{rr}\right)$ is equal to $r/(2t)$.
This implies that the time variable $t$ used in the calculation above
is in fact $\tilde t$ in the notations of Sec.~\ref{ssec:filament}.
It is thus straightforward to convert the present formula for the
self-similar solution $\vec M(r,t)$ to the convention $m=1$ by the
replacement $t\to t/2$, as was done in Fig.~\ref{fig:SS}.

\bibliography{xxz_lle_paper.bib}{}
\end{document}